\documentclass[a4paper,12pt]{article}

\usepackage{amssymb,latexsym}
\usepackage{amsmath}
\usepackage{bbm}
\usepackage{latexsym}
\usepackage{epsfig}
\usepackage{verbatim} 

\usepackage{fullpage}

\makeatletter \@addtoreset{equation}{section} 
\makeatother

\begin{document} 
\begin{titlepage}
	\thispagestyle{empty} 
	
	\begin{flushright}

	\end{flushright}
	
	\vspace{35pt} 
	\begin{center}
		{ \Large{\bf Conformal chiral boson models on twisted doubled \\ \vspace{4mm}
		 tori
		and non-geometric string vacua
		 }}
		
		\vspace{35pt}
		
		\bf{Spyros D. Avramis, Jean-Pierre Derendinger
		 and Nikolaos Prezas}
		
		\vspace{15pt}
		
		\footnotesize{\tt avramis, derendinger, prezas@itp.unibe.ch}
		
		\vspace{35pt}
		
		{\it   Albert Einstein Center for Fundamental Physics\\
		Institute for Theoretical Physics, University of Bern\\
		Sidlerstrasse 5, CH-3012 Bern, Switzerland
		}
				
		\vspace{50pt}
		
		{ABSTRACT}
	\end{center}
	
\noindent
We derive and analyze the conditions for quantum conformal and Lorentz invariance of the duality symmetric interacting chiral boson sigma-models, which are conjectured to describe non-geometric string theory backgrounds. The one-loop Weyl and Lorentz anomalies are computed for the general case using the background field method. Subsequently, our results are applied to a class of (on-shell) Lorentz invariant 
chiral boson models which are based on twisted doubled tori. Our findings are in agreement with those expected from the effective supergravity approach, thereby 
firmly establishing that the chiral boson models under consideration provide the string worldsheet description of ${\cal N}=4$ gauged supergravities with electric gaugings.
Furthermore, they demonstrate  that
twisted doubled tori are indeed the doubled internal geometries underlying a large class of non-geometric string compactifications. For compact gaugings the associated chiral boson models are automatically conformal,  a fact that is explained by showing that they are actually chiral WZW models in disguise.

	\vspace{10pt}

	\end{titlepage}

\baselineskip 6 mm
 
\tableofcontents

\section{Introduction}

 It has been realized during recent years that string dualities allow for an extension of usual string backgrounds, namely those comprising of a well-defined geometry and physical fluxes, to a broader class that has been dubbed
  {\it non-geometric} (see \cite{Wecht:2007wu} for a review). 
The latter might involve geometries and/or flux configurations that are ill-defined from the perspective of a point particle. The existence of such backgrounds can be inferred from two complementary approaches.

The first approach is indirect and is based on the effective supergravities that describe string compactifications after integrating out modes originating from the compact internal dimensions. A crucial feature of these supergravity theories is the existence of global non-compact duality symmetries \cite{Cremmer:1977tt, Cremmer:1979up}. Promoting a part of this duality symmetry to a local gauge symmetry yields a gauged supergravity theory which contains a potential for the various scalar fields. From the higher-dimensional viewpoint, a subset of such gaugings is known to correspond to compactifications with non-trivial internal geometries and/or 
 internal fluxes. However, most of these gaugings do not admit such a lift; instead, there is convincing evidence that their higher-dimensional origin is in terms of a non-geometric background.

The description of the most general gauged supergravity compatible with a given number of supersymmetries is based on an object known as the embedding tensor \cite{deWit:2005ub}.  The embedding tensor provides a duality-covariant formulation of the theory by characterizing the gauge algebra through its embedding in the duality group and, in cases with many supersymmetries, completely determines the theory. The study of the embedding tensor and the subsequent classification of gauged supergravities has revealed that most theories obtained by requiring duality covariance result from compactifications of string theory (or M--theory) on non-geometric backgrounds \cite{Shelton:2005cf, Aldazabal:2006up, Shelton:2006fd}.

The second approach to non-geometric backgrounds stems directly from the various string dualities and in particular from T-duality. The latter is an inherently stringy symmetry as it exchanges momentum modes with the winding modes that originate from the extended nature of the string. The simplest examples of non-geometric backgrounds, known as T-folds and which will be the focal point of this paper, feature transition functions between patches of the internal space that contain T-duality transformations \cite{Hellerman:2002ax, Dabholkar:2002sy, Kachru:2002sk}; this renders string propagation possible on various ill-defined geometries and/or configurations of fluxes. In order to analyze 
such non-geometric configurations, one must therefore employ a T-duality covariant formulation which treats momentum and winding modes in a democratic fashion. 

A natural way to implement this idea is by doubling the dimensionality of the compact internal space, with the extra dimensions being conjugate to the winding modes of the string 
\cite{Witten:1988zd, Duff:1989tf, Tseytlin:1990nb, Tseytlin:1990va, Maharana:1992my}. 
More recently, Hull advocated a geometric description of non-geometric backgrounds, in particular of T-folds, in terms of doubled tori where the action of the T-duality group is manifest \cite{Hull:2004in, Hull:2006va, Hull:2007zu}. Such constructions naturally raise the question whether these doubled geometries are just a convenient bookkeeping device or have a deeper physical significance. 

A first step towards answering this question was taken in the context of the effective supergravity approach. It was suggested in \cite{Dall'Agata:2007sr} that twisting the doubled torus, {\it i.e.}~promoting it to a local group manifold, yields the underlying geometries of a class of
${\cal N}=4$ gauged supergravities.\footnote{A similar idea was presented earlier in \cite{Hull:2007jy} but only for a very specific class of gauge algebras.} This twisted doubled torus (TDT) is actually the group manifold corresponding to the supergravity gauge algebra
subject to global identifications.
In the lack of an explicit theory that is defined on the doubled geometry and could have been directly reduced to the effective supergravity, the evidence presented in  \cite{Dall'Agata:2007sr}  was based on a comparison of the potential resulting from a Scherk--Schwarz type reduction on an ordinary twisted torus \cite{Scherk:1979zr, Kaloper:1999yr} with the generic form of the potential in gauged supergravity. The main result of ref.~\cite{Dall'Agata:2007sr} was 
a novel interpretation of the embedding tensor as being 
geometric flux but in the doubled torus.

More recently, further evidence that the TDT can be serious candidates for actual string backgrounds came from the complementary worldsheet approach. From the worldsheet viewpoint, treating momentum and winding modes on an equal footing amounts to treating the left- and right-moving modes of a closed string independently, {\it i.e.}~requires a chiral boson theory in two dimensions. Such theories have a long history starting with \cite{Siegel:1983es} and their formulation presented in \cite{Floreanini:1987as} was the starting point for constructing an interacting chiral boson model for closed strings \cite{Tseytlin:1990nb, Tseytlin:1990va}. These theories are manifestly duality invariant, at the expense of lost worldsheet Lorentz invariance. Restoring Lorentz invariance on shell imposes stringent conditions on the background fields, whose known solutions were very few. Therefore it is remarkable that, as shown in \cite{Dall'Agata:2008qz}, chiral boson models based on twisted doubled tori and supplemented by a flux for the Wess--Zumino term turn out to be Lorentz invariant.\footnote{Worldsheet models for T-folds along with ideas similar to the twisted doubled tori have also been pursued in refs.~\cite{Albertsson:2008gq, Hull:2009sg, ReidEdwards:2009nu}. These models are second-order formulations and therefore require explicit constraints in order to reduce the number of propagating fields by half, in contrast to the first-order models of refs.~\cite{Tseytlin:1990nb, Tseytlin:1990va, Dall'Agata:2008qz}. Another first-order model for T-folds has been proposed in \cite{Halmagyi:2009te}. Although the exact relationship between all these models remains unclear for genuine non-geometric backgrounds, we believe that the models of refs.~\cite{Tseytlin:1990nb, Tseytlin:1990va} and  in particular the Lorentz invariant class of \cite{Dall'Agata:2008qz} corresponding to the TDT, provide the cleanest route towards analyzing quantum effects.}

The goal of the present paper is to examine whether the classically Lorentz and Weyl invariant models of ref.~\cite{Dall'Agata:2008qz} satisfy the requirement of conformal invariance at the quantum level, as required for worldsheet theories underlying (non-geometric) string compactifications. To this end, we compute the one-loop effective action of the general model of \cite{Tseytlin:1990nb, Tseytlin:1990va} and then we focus on the Lorentz invariant class of \cite{Dall'Agata:2008qz}. Earlier work on the calculations of beta functionals in the interacting chiral boson models has been carried out in \cite{Berman:2007xn, Berman:2007yf} for the case where the background fields depend trivially on the doubled coordinates (but non-trivially on the non-compact spacetime coordinates). Here, we consider fields with arbitrary dependence on the full doubled geometry, which is necessary in order to obtain effective spacetime theories corresponding to generic non-geometric compactifications. 

Specializing to the TDT case, the vanishing of the Weyl anomaly yields conditions in order for the models to provide actual string vacua. These conditions are the same as those required to minimize 
 the corresponding ${\cal N}=4$
 supergravity potential, thereby  establishing the connection of these models to gauged supergravity. In addition, the Weyl anomaly is identically zero
for compact gaugings and we explain this fact by showing that the  corresponding models are 
 actually chiral Wess--Zumino--Witten (WZW) models. 

This paper is organized as follows. In Section \ref{icbmodel} we review the interacting chiral boson model and we state the conditions for on-shell Lorentz invariance and their known solutions. In Section \ref{effectiveaction} we employ the background field method to obtain the one-loop effective action for the model and we determine the Weyl and Lorentz anomalies. In Section \ref{conformaltdt} we first review worldsheet TDT models and the conditions for classical Lorentz invariance. Then we compute the Weyl and Lorentz anomalies and we relate the conformal invariance condition to the minimization of the potential of the associated gauged supergravity. Subsequently, we focus on compact gaugings and we demonstrate that they correspond to WZW models. Finally,
we discuss some illustrative examples and, in Section 5, we present some
 possible directions for future research. Our notation and our differential geometry conventions are summarized in the Appendix.

\section{The interacting chiral boson model}
\label{icbmodel}

We start with the general duality-invariant sigma-model of interacting
chiral bosons which was proposed by Tseytlin \cite{Tseytlin:1990nb, Tseytlin:1990va} as a natural generalization of the Floreanini--Jackiw (FJ) Lagrangian \cite{Floreanini:1987as}.
This is a two-dimensional theory described by the action 
\begin{equation}
S=\frac{1}{2} \int d^2\sigma \Big( - H_{IJ}({\mathbb Y}) \partial_1 {\mathbb Y}^I \partial_1 {\mathbb Y}^J+
\big(\eta_{IJ}({\mathbb Y})+C_{IJ}({\mathbb Y})\big) \partial_0 {\mathbb Y}^I \partial_1 {\mathbb Y}^J\Big),
\label{icb}
\end{equation}
where we have  $2d$ scalars  ${\mathbb Y}^I, I=1,\ldots,2d$ and the background fields $H_{IJ}$ and $\eta_{IJ}$ are symmetric while $C_{IJ}$ is antisymmetric. The indices $I,J,\ldots$ can be separated into two sets reflecting the ``doubling'' ${\mathbb Y}^I = \{ y^i, y^{\tilde i}\}\equiv \{y^i, \tilde y_i\}$ with $i=1, \ldots,d$.

The free theory corresponds to
\begin{equation}
H_{IJ}= \left( 
	\begin{array}{cc}
		\mathbbm{1}_d & 0 \\
		0 & \mathbbm{1}_d 
	\end{array}
	\right)\ , \quad
	\eta_{IJ}= \left( 
	\begin{array}{cc}
		0 & \mathbbm{1}_d  \\
		 \mathbbm{1}_d & 0  
	\end{array}
	\right)\ , \quad C_{IJ} = 0\ .
	\label{freecbm}
\end{equation}	
Then, eq.~(\ref{icb}) boils down to two sets of FJ Lagrangians for $d$ chiral and $d$ antichiral bosons $y^i \pm \tilde y_i$ which can be recombined to the usual two-dimensional theory for $d$ free bosons $y^i$.

The equations of motion following from the above action are  
\begin{equation}
2 \partial_1 V_I + \partial_I H_{JK}\partial_1 {\mathbb Y}^J \partial_1 {\mathbb Y}^K - G_{IJK} \partial_0 {\mathbb Y}^J \partial_1 {\mathbb Y}^K - 2 \eta_{JL} \Gamma^L_{IK}(\eta) \partial_0 {\mathbb Y}^J \partial_1 {\mathbb Y}^K=0\ , 
\label{eom} 
\end{equation}
where $\Gamma^K_{IJ}(\eta)$ are the Christoffel symbols constructed out of $\eta_{IJ}$, the ``generalized three-form flux'' $G_{IJK}$ is given by
\begin{equation}
G_{IJK} = \partial_I C_{JK} + \partial_J C_{KI} + \partial_K C_{IJ}\ ,
\end{equation}  
and $V_I$ is defined as
\begin{equation}
V_I \equiv \eta_{IJ} \partial_0 {\mathbb Y}^J - H_{IJ}\partial_1 {\mathbb Y}^J\ .
\label{Vdef} 
\end{equation}

It is interesting to examine the symmetries of the above theory. First, we note that (\ref{icb}) is the gauge-fixed form of a more general action, given by
\begin{equation}
S=\frac{1}{2} \int d^2 \sigma\, e \Big( - H_{IJ}({\mathbb Y}) \nabla_1 {\mathbb Y}^I \nabla_1 {\mathbb Y}^J+
\big(\eta_{IJ}({\mathbb Y})+C_{IJ}({\mathbb Y})\big) \nabla_0 {\mathbb Y}^I \nabla_1 {\mathbb Y}^J\Big)\ ,
\label{icbcov}
\end{equation}
where $e^\alpha_\mu$ is the zweibein and $\nabla_\alpha = e^\mu_\alpha \partial_\mu$ is the associated worldsheet covariant derivative. From eq.~(\ref{icbcov}), it is obvious that the theory is manifestly invariant under classical Weyl transformations and diffeomorphisms. However, since the time and space coordinates are treated on a different footing, the theory is not manifestly Lorentz invariant. Indeed, the standard condition for  Lorentz invariance, $\epsilon^{\mu\nu} T_{\mu\nu}=0$, is not identically satisfied but leads to 
\begin{equation}
\eta_{IJ} \left(\partial_0 {\mathbb Y}^I \partial_0 {\mathbb Y}^J + \partial_1 {\mathbb Y}^I \partial_1 {\mathbb Y}^J\right) - 2 H_{IJ} \partial_0 {\mathbb Y}^I \partial_1 {\mathbb Y}^J = 0\ . 
\label{lorcon}
\end{equation}
This constraint can be recast \cite{Tseytlin:1990va} in the more suggestive form 
\begin{equation}
\left(\eta- H\eta^{-1}H\right)_{IJ} \partial_1 {\mathbb Y}^I \partial_1 {\mathbb Y}^J + \eta^{IJ} V_I V_J = 0\ ,
\label{lorentzconstraint} 
\end{equation}
where $V_I$ is the vector introduced in (\ref{Vdef}) and $\eta^{IJ}$ is the inverse of $\eta_{IJ}$. Hence, if the following equations hold 
\begin{equation}
\eta = H \eta^{-1} H\ ,  
\label{lorentz1}
\end{equation}
and 
\begin{equation}
\eta^{IJ} V_I V_J = 0\ , 
\label{lorentz2}
\end{equation}
then classical Lorentz invariance is restored. These two conditions
are actually sufficient
but not necessary to verify eq.~(\ref{lorcon}). Nevertheless, the models
which are known to satisfy condition (\ref{lorcon}) actually verify eqs.~(\ref{lorentz1})
and (\ref{lorentz2}) independently.

The first condition, eq.~(\ref{lorentz1}), is easy to satisfy by appropriately choosing $H_{IJ}$ and $\eta_{IJ}$. It is the second condition, eq.~(\ref{lorentz2}), that poses non-trivial constraints on $H_{IJ}$, $\eta_{IJ}$ and $C_{IJ}$ and seriously restricts the possible Lorentz invariant theories.\footnote{Although eq.~(\ref{lorentz2}) involves only $H_{IJ}$ and $\eta_{IJ}$, the fact that it should be satisfied on shell imposes restrictions on $C_{IJ}$ through the equations of motion (\ref{eom}).} The only known classes of solutions of eqs.~(\ref{lorentz1}) and (\ref{lorentz2}) are the following:

\vskip .2cm
\noindent $\bullet$ {\bf Standard sigma-models}. These are theories where the background
fields take the form
\begin{equation}
\label{stansin}
H_{IJ} = \left( 
\begin{array}{cc}
G_{ij} - B_{ik} G^{kl} B_{lj} & B_{ik} G^{kj}\\
- G^{ik} B_{kj} & G^{ij} 
\end{array}
\right)\ , \quad 
\eta_{IJ} = \left( 
\begin{array}{cc}
0 & \mathbbm{1}_d\\
\mathbbm{1}_d & 0 
\end{array}
\right)\ , \quad C_{IJ}={\rm const}.\ , 
\end{equation}
with $G_{ij}=G_{ji}$ and $B_{ij}=-B_{ji}$ being either constant or dependent only on $y^i$, 
{\it i.e.}~on half of the doubled coordinates.
These solutions were first found in \cite{Tseytlin:1990va} and they are equivalent to a standard 
sigma-model for half of the
doubled coordinates with $G_{ij}$ serving as a background metric and $B_{ij}$ as a two-form potential, {\it i.e.}
\begin{equation}
S = \int d^2\sigma ( g^{\mu \nu} G_{ij} + \epsilon^{\mu\nu} B_{ij} ) \partial_\mu y^i \partial_\nu y^j \ .
\end{equation}
\vskip .2cm
\noindent $\bullet$ {\bf Twisted doubled tori}. These backgrounds describe $2d$-dimensional group manifolds ${\cal G}$ subject to discrete identifications. They were introduced in ref.~\cite{Dall'Agata:2007sr} with the objective of providing a unified description of gauged supergravities arising from geometric and non-geometric string theory compactifications. In \cite{Dall'Agata:2008qz}, it was proven that these backgrounds satisfy the classical Lorentz invariance constraints if a precise relation exists between the generalized flux $G_{IJK}$ and the structure constants of ${\cal G}$. 
For the particular case of compact gaugings, these models, as anticipated
in \cite{Dall'Agata:2008qz}, will be shown to be chiral WZW models in disguise.
We will elaborate more on these backgrounds and the corresponding chiral boson models in Section 4.
\vskip .2cm
\noindent $\bullet$ {\bf Interacting non-abelian chiral scalars}. These models were originally constructed in \cite{Depireux:1988yi, Gates:1987sy} by bosonizing a particular non-abelian massless Thirring model and they were further discussed in \cite{Tseytlin:1990va}. Although they include chiral WZW models as special cases, the relevance of the generic model of this type to the considerations of this paper is unclear.
\vskip .2cm
\noindent $\bullet$ Theories where $H_{IJ}$ and $\eta_{IJ}$ are  subject to the condition $\eta_{IJ} = \pm H_{IJ}$ but otherwise arbitrary. The fact that such theories solve the classical Lorentz invariance constraints was proven in \cite{Dall'Agata:2008qz}. However, as the associated sigma-models describe $2d$ bosons of equal chirality, these theories suffer from a quantum Lorentz anomaly and are of limited interest.
\vskip .2cm

Given the above classically consistent theories, one is still faced with the task of examining whether this consistency holds at the quantum level. This amounts to computing the relevant contributions to the quantum effective action which include the standard Weyl anomaly as well as the global Lorentz anomaly, the latter being due to the fact that the models under consideration contain chiral bosons. Only backgrounds for which the Weyl and Lorentz anomalies vanish can be consistent string vacua.

\section{General computation of the effective action}
\label{effectiveaction}

Having presented the classical theory in sufficient detail, we may now calculate the quantum effective action and extract the contributions corresponding to the Weyl and Lorentz anomalies. That calculation is based on the standard background field method \cite{AlvarezGaume:1981hn, Braaten:1985is, Mukhi:1985vy, Howe:1986vm} appropriately adapted to the sigma-model under consideration. Our strategy will be similar to that used in \cite{Berman:2007xn}, but the arrangement of the various terms in the expansion will be different. Our final result will be a master expression which can be used to obtain the Weyl and Lorentz anomalies for any particular case of interest.

\subsection{Background field expansion}

The starting point for the application of the background field method to the standard sigma-model is the expansion of the fields ${\mathbb Y}^I$ according to ${\mathbb Y}^I={\mathbb Y}^I_{cl}+\pi^I$, where ${\mathbb Y}^I_{cl}$ is a solution of the classical equations of motion and $\pi^I$ is the fluctuation. Since the fluctuation $\pi^I$ does not transform as a vector and hence yields a non-covariant expansion of the action, it is more convenient to trade it for the field $\xi^I$, defined as the tangent to the geodesic from ${\mathbb Y}^I_{cl}$ to ${\mathbb Y}^I_{cl}+\pi^I$ whose length equals the arc length of the geodesic; this field obviously transforms as a vector and the resulting expansion is covariant. It turns out that consecutive terms in the expansion of the action can be represented in a concise form \cite{Mukhi:1985vy} by the following relation
\begin{equation}
S_n = {1 \over n!} {\cal D}^n S \equiv {1 \over n!} \left( \int d^2 \sigma \xi^I(\sigma) {D \over D {\mathbb Y}^I (\sigma)} \right)^n S\ ,
\label{mukhi}
\end{equation}
where ${\mathbb Y}^I$ will henceforth stand for ${\mathbb Y}^I_{cl}$ and where $D / D {\mathbb Y}^I (\sigma)$ is the covariant functional derivative with respect to ${\mathbb Y}^I (\sigma)$. The merits of using this method are that (a) the action of the operator ${\cal D}$ on the various objects appearing on the expansion of the action is particularly simple \cite{Mukhi:1985vy} and (b) the formula (\ref{mukhi}) leads to a simple recursive algorithm determining $S_n$ in terms of $S_{n-1}$.

The method just described can be applied in a straightforward manner to the chiral sigma-model under consideration, the only particularity being that now we have two objects playing the role of the ``metric'', namely $\eta_{IJ}$ and $H_{IJ}$. Hence one has the option to define $\xi^I$ either in terms of geodesics of $\eta_{IJ}$ or in terms of geodesics of $H_{IJ}$, with the resulting expansion involving covariant derivatives and tensors with respect to the chosen metric. Although in the present paper we will use the $\eta$-covariant form of the expansion, each of these expansions is potentially useful for certain applications and therefore we will present both of them for the sake of completeness. 

\vskip .2cm
\noindent $\bullet$ {\bf The \boldmath{$\eta$}-covariant expansion.} We start by considering the covariant expansion with respect to $\eta_{IJ}$. In this case, the first order terms in $\xi^I$ are found by acting once with the operator ${\cal D}$ of (\ref{mukhi}) on $S$. The result is
\begin{eqnarray}
S_1 &=& \int d^2 \sigma \Big( \frac{1}{2} \eta_{IJ} (\partial_0 {\mathbb Y}^I D_1 \xi^J + D_0 \xi^I \partial_1 {\mathbb Y}^J) - H_{IJ} \partial_1 {\mathbb Y}^I D_1 \xi^J  \nonumber\\ 
&& \qquad\qquad - \frac{1}{2} D_K H_{IJ} \xi^K \partial_1 {\mathbb Y}^I \partial_1 {\mathbb Y}^J + \frac{1}{2} G_{IJK} \xi^K \partial_0 {\mathbb Y}^I \partial_1 {\mathbb Y}^J \Big)\
 \label{firstordetacov}
\end{eqnarray}
and is easily seen to vanish on the equations of motion (\ref{eom}), as it should. Acting on (\ref{firstordetacov}) with ${\cal D}$ and including a factor of $1/2$, we find that the second-order terms read
\begin{eqnarray}
\!\!\!\!S_2 &=& \frac{1}{2}\int d^2 \sigma \bigg(- H_{IJ} D_1 \xi^I D_1 \xi^J + \eta_{IJ} D_0\xi^I D_1 \xi^J \nonumber\\
&&\qquad\qquad + \left( {1 \over 2} D_J G_{IKL} + R_{KIJL} \right) \xi^I \xi^J \partial_0 {\mathbb Y}^K \partial_1 {\mathbb Y}^L \nonumber\\
&&\qquad\qquad - \frac{1}{2}( D_I D_J H_{KL} + H_{KM} R^{M}{}_{IJL} + H_{LM} R^{M}{}_{IJK} ) \xi^I \xi^J \partial_1 {\mathbb Y}^K \partial_1 {\mathbb Y}^L \nonumber\\
&&\qquad\qquad + {1 \over 2} G_{IJK} \xi^K (\partial_0 {\mathbb Y}^I D_1 \xi^J + D_0 \xi^I \partial_1 {\mathbb Y}^J) - 2 D_K H_{IJ} \xi^K D_1 \xi^I \partial_1 {\mathbb Y}^J
 \bigg)\ .
 \label{secordetacov}
\end{eqnarray}
In the above expressions, the covariant derivatives and the Riemann tensor are constructed out of $\eta_{IJ}$. 
\vskip .2cm
\noindent $\bullet$ {\bf \boldmath{$H$}-covariant expansion.} In the covariant expansion with respect to $H_{IJ}$, the first-order terms in $\xi^I$ are given by
\begin{eqnarray}
S_1 &=& \int d^2 \sigma \Big( -H_{IJ} \partial_1 {\mathbb Y}^I D_1 \xi^J + \frac{1}{2} ( D_K \eta_{IJ} + G_{IJK} ) \xi^K \partial_0 {\mathbb Y}^I \partial_1 {\mathbb Y}^J \nonumber\\ 
&& \qquad\qquad +\frac{1}{2}
\eta_{IJ} ( \partial_0 {\mathbb Y}^I D_1 \xi^J + D_0 \xi^I \partial_1 {\mathbb Y}^J )\Big)
 \label{firstordHcov}
\end{eqnarray}
and again vanish on the equations of motion. The second-order terms read
\begin{eqnarray}
S_2 &=& \frac{1}{2}\int d^2 \sigma \bigg(- H_{IJ} D_1 \xi^I D_1 \xi^J +  \eta_{IJ} D_0\xi^I D_1 \xi^J - R_{KIJL} \xi^I \xi^J \partial_1 {\mathbb Y}^K \partial_1 {\mathbb Y}^L\nonumber\\
&&\qquad\qquad +\frac{1}{2}(D_J G_{IKL} + D_I D_J \eta_{KL} + \eta_{KM} R^{M}{}_{IJL}
 + \eta_{LM} R^{M}{}_{IJK} ) \xi^I \xi^J \partial_0 {\mathbb Y}^K \partial_1 {\mathbb Y}^L \nonumber\\
&&\qquad\qquad + \left( D_K \eta_{IJ} + {1 \over 2} G_{IJK} \right) \xi^K (\partial_0 {\mathbb Y}^I D_1 \xi^J + D_0 \xi^I \partial_1 {\mathbb Y}^J)
 \bigg)\ .
 \label{secordHcov}
\end{eqnarray}
Now, the covariant derivatives and the Riemann tensor are constructed out of $H_{IJ}$.
\vskip .2cm
 
At this point, a comment is in order. When writing the second-order action, one has the choice of invoking the classical equations of motion (\ref{eom}) (or, equivalently, the vanishing of the first-order action (\ref{firstordetacov}) or (\ref{firstordHcov})) to rearrange  various terms. Although this choice may make certain cancellations of terms manifest, the resulting second-order action is no longer expressed in terms of covariant derivatives (see for example \cite{Berman:2007xn}) and in our case yields a rather complicated form for the effective action. For that reason, we will refrain from using the classical equations of motion at this point, reserving the option to apply them at a later stage, if necessary.

\subsection{Structure of the effective action}

Given the above expansion, we can determine the form of the one-loop effective action obtained after integrating out the $\xi$-fluctuations. The effective action is given by $S_{{\rm eff}}=S_{{\rm cl}}+\Gamma$ where $\Gamma$ represents the one-loop corrections and is given by the standard formula
\begin{equation}
\exp\left( {\rm i} \Gamma[{\mathbb Y}] \right) = \int {\cal D} \xi \exp\left( {\rm i} S_2 [{\mathbb Y};\xi] \right)\ .
\end{equation}
Decomposing $S_2$ into ``kinetic'' and ``interacting'' parts, $S_2=S_{2,{\rm k}}+S_{2,{\rm i}}$ and expanding $e^{ {\rm i} S_{2,{\rm i}}}$, we obtain\footnote{We normalize the free determinant to unity.}
\begin{equation}
e^{ {\rm i} \Gamma} = \int {\cal D} \xi \left( 1 + {\rm i} S_{2,{\rm i}} - {1 \over 2} S_{2,{\rm i}}^2 + \ldots \right) e^{ {\rm i} S_{2,{\rm k}} } = 1 + {\rm i} \left\langle S_{2,{\rm i}} \right\rangle - {1 \over 2} \left\langle S_{2,{\rm i}}^2 \right\rangle + \ldots\ ,
\end{equation}
where $\langle \cdot \rangle$ denotes the expectation value with respect to $S_{2,{\rm k}}$. Hence
\begin{equation}
\Gamma = \left\langle S_{2,{\rm i}} \right\rangle + {{\rm i} \over 2} \left\langle S_{2,{\rm i}}^2 \right\rangle_{\rm conn} + \ldots\ .
\label{eff1}
\end{equation}
For the calculation of the Weyl (and Lorentz) anomaly we only need the divergent contributions to the above expression. These originate from the terms given in eq.~(\ref{eff1}) and, more specifically, they arise \cite{Braaten:1985is} from the $\langle \xi\xi \rangle$ single contractions in the first term of (\ref{eff1}) and the $\langle \xi \partial \xi \xi \partial \xi \rangle$ double contractions in the second term of (\ref{eff1}).

The calculation of the above contractions using the quadratic form $- H_{IJ} \partial_1 \xi^I \partial_1 \xi^J + \eta_{IJ} \partial_0 \xi^I \partial_1 \xi^J$ as our kinetic term is impossible due to the non-trivial dependence of $H_{IJ}$ and $\eta_{IJ}$ on ${\mathbb Y}^I$ for general background field configurations.
It is clear that one can introduce a vielbein $E^A_I$ that diagonalizes $H_{IJ}$ according to 
\begin{equation}
H_{IJ} = H_{AB} E^A_I E^B_J\ ;\qquad H_{AB} = \left(
\begin{array}{cc} 
\mathbbm{1}_d & 0  \\
0 & \mathbbm{1} _d 
\end{array} \right)
\label{vb1}
\end{equation}
so that $- H_{IJ} \partial_1 \xi^I \partial_1 \xi^J$ gives rise to a canonical kinetic term $-H_{AB} \partial_1 \xi^A \partial_1 \xi^B$ as well as terms involving the derivatives of the
vielbein.

What is perhaps less obvious is that there exists a vielbein that simultaneously satisfies (\ref{vb1}) and also
\begin{equation}
\eta_{IJ} = \eta_{AB} E^A_I E^B_J\ ;\qquad \eta_{AB} = \left(
\begin{array}{cc} 
\mathbbm{1}_d & 0  \\
0 & - \mathbbm{1}_d  
\end{array} \right)\ .
\label{vb2}
\end{equation}
To see this, we recall that the relation (\ref{vb1}) still entails an ambiguity up to ${\rm O}(2 d)$ rotations $E^A_I \to R^A_{\phantom{A} B} E^B_I$. Under such rotations, the symmetric matrix $\eta_{AB}$, as defined by the first of (\ref{vb2}), transforms as $\eta_{AB} \to R^C_{\phantom{C} A} \eta_{CD} R^D_{\phantom{D} B}$ and can thus be brought to a diagonal form by a particular choice of $R$. Now, writing the Lorentz invariance condition (\ref{lorentz1}) on the tangent space, we obtain $\eta_{AC} \delta^{CD} \eta_{DB} = \delta_{AB}$, which implies that the diagonal form of $\eta_{AB}$ can have only $\pm 1$ entries. Finally, since the vanishing of the quantum Lorentz anomaly demands that the sum of these entries is zero \cite{Tseytlin:1990va}, $\eta_{AB}$ can take the form given in  (\ref{vb2}). Notice
that bringing $H_{IJ}$ and $\eta_{IJ}$ to canonical form still leaves a residual
${\rm O}(d) \times {\rm O}(d)$ tangent space symmetry.

Using the vielbein basis, we can decompose the action  (\ref{secordetacov}) or
(\ref{secordHcov}) into kinetic and interaction terms for the tangent-space fields $\xi^A$. The kinetic term reads
\begin{equation}
S_{2,{\rm k}} = \frac{1}{2}\int d^2 \sigma \left( - H_{AB} \partial_1 \xi^B \partial_1 \xi^B +  \eta_{AB} \partial_0\xi^A \partial_1 \xi^B \right) \ ,
 \label{kinetic}
\end{equation}
and, as mentioned in \S\ref{icbmodel}, it is just the sum of Floreanini--Jackiw actions for $d$ chiral and $d$ antichiral bosons in the chiral basis. As for the interaction terms, they can be written in the schematic form
\begin{equation}
S_{2,{\rm i}} = \frac{1}{2} \int d^2 \sigma ( {\cal S}_{AB} \xi^A \xi^B + {\cal Q}_{AB} \xi^A \partial_1 \xi^B + {\cal P}_{AB} \xi^A \partial_0 \xi^B  )\ .
\label{interaction}
\end{equation}
Inserting (\ref{interaction}) in the expression (\ref{eff1}) for the effective action and keeping only the relevant terms, we obtain the expression
\begin{eqnarray}
\Gamma &=& {1 \over 2} \int d^2 \sigma \Big( {\cal S}_{AB} \langle\langle \xi^A \xi^B \rangle\rangle + {1 \over 4} {\cal Q}_{AB} {\cal Q}_{CD} \langle\langle \xi^A \partial_1 \xi^B \xi^C \partial_1 \xi^D \rangle\rangle \nonumber\\ 
&& \qquad\qquad + {1 \over 2} {\cal Q}_{AB} {\cal P}_{CD} \langle\langle \xi^A \partial_1 \xi^B \xi^C \partial_0 \xi^D \rangle\rangle \nonumber\\ 
&& \qquad\qquad + {1 \over 4} {\cal P}_{AB} {\cal P}_{CD} \langle\langle \xi^A \partial_0 \xi^B \xi^C \partial_0 \xi^D \rangle\rangle \Big)
\label{eff2}
\end{eqnarray}
where we introduced the shorthands
\begin{equation}
\langle\langle \xi^A \xi^B \rangle\rangle \equiv \langle \xi^A(\sigma) \xi^B(\sigma) \rangle
\label{singlecont}
\end{equation}
and
\begin{equation}
\langle\langle \xi^A \partial_\mu \xi^B \xi^C \partial_\nu \xi^D \rangle\rangle \equiv {\rm i} \int d^2 \sigma^\prime \langle \xi^A(\sigma) \partial_\mu \xi^B(\sigma) \xi^C (\sigma^\prime) \partial^\prime_\nu \xi^D(\sigma^\prime) \rangle\ . 
\label{doublecont}
\end{equation}
Now, by inspection of the action (\ref{secordetacov}) or (\ref{secordHcov}), one easily sees that ${\cal S}_{AB}$, ${\cal Q}_{AB}$ and ${\cal P}_{AB}$ have the form
\begin{eqnarray}
\label{sqp}
{\cal S}_{AB} &=& {\cal S}^{11}_{AB,IJ} \partial_1 {\mathbb Y}^I \partial_1 {\mathbb Y}^J + {\cal S}^{01}_{AB,IJ} \partial_0 {\mathbb Y}^I \partial_1 {\mathbb Y}^J\ , \nonumber\\
{\cal Q}_{AB} &=& {\cal Q}^{1}_{AB,I} \partial_1 {\mathbb Y}^I + {\cal Q}^{0}_{AB,I} \partial_0 {\mathbb Y}^I\ , \\
{\cal P}_{AB} &=& {\cal P}^{1}_{AB,I} \partial_1 {\mathbb Y}^I\ ,\nonumber
 \end{eqnarray}
where the various contributions will be explicitly given below. 
Inserting (\ref{sqp}) in (\ref{eff2}), we can write our final expression for the effective action in the form
\begin{equation}
\Gamma = {1 \over 2} \int d^2 \sigma \left( \Gamma^{00}_{IJ} \partial_0 {\mathbb Y}^I \partial_0 {\mathbb Y}^J + \Gamma^{01}_{IJ} \partial_0 {\mathbb Y}^I \partial_1 {\mathbb Y}^J + \Gamma^{11}_{IJ} \partial_1 {\mathbb Y}^I \partial_1 {\mathbb Y}^J \right)
\label{eff3}
\end{equation}
with the three terms given by
\begin{eqnarray}
\Gamma^{00}_{IJ} &=& {1 \over 4} {\cal Q}^{0}_{AB,I} {\cal Q}^{0}_{CD,J} \langle\langle \xi^A \partial_1 \xi^B \xi^C \partial_1 \xi^D \rangle\rangle\ , \nonumber\\
\Gamma^{01}_{IJ} &=& {\cal S}^{01}_{AB,IJ} \langle\langle \xi^A \xi^B \rangle\rangle + {1 \over 2} {\cal Q}^{0}_{AB,I} {\cal Q}^{1}_{CD,J} \langle\langle \xi^A \partial_1 \xi^B \xi^C \partial_1 \xi^D \rangle\rangle\ \nonumber\\
&+& {1 \over 4} \left( {\cal Q}^{0}_{AB,I} {\cal P}^{1}_{CD,J} + {\cal P}^{1}_{AB,I} {\cal Q}^{0}_{CD,J} \right)  \langle\langle \xi^A \partial_1 \xi^B \xi^C \partial_0 \xi^D \rangle\rangle\ , \nonumber\\
\Gamma^{11}_{IJ} &=& {\cal S}^{11}_{AB,IJ} \langle\langle \xi^A \xi^B \rangle\rangle + {1 \over 4} {\cal Q}^{1}_{AB,I} {\cal Q}^{1}_{CD,J} \langle\langle \xi^A \partial_1 \xi^B \xi^C \partial_1 \xi^D \rangle\rangle\ \nonumber\\
&+& {1 \over 4} \left( {\cal Q}^{1}_{AB,I} {\cal P}^{1}_{CD,J} + {\cal P}^{1}_{AB,I} {\cal Q}^{1}_{CD,J} \right) \langle\langle \xi^A \partial_1 \xi^B \xi^C \partial_0 \xi^D \rangle\rangle \nonumber\\
&+& {1 \over 4} {\cal P}^{1}_{AB,I} {\cal P}^{1}_{CD,J} \langle\langle \xi^A \partial_0 \xi^B \xi^C \partial_0 \xi^D \rangle\rangle\  .
\label{eff2terms}
\end{eqnarray}
To put the above expression to use, we must calculate the contractions appearing in this expression and we must write down explicit formulas for the quantities ${\cal S}$, ${\cal Q}$ and ${\cal P}$. This will be the objective of the following two subsections.

\subsection{Propagators and contractions}

To calculate the contractions appearing in (\ref{eff2terms}) we first have to obtain the boson propagator corresponding to the ``kinetic'' Lagrangian (\ref{kinetic}). Using the diagonal form of the $H$ and $\eta$ matrices, one easily finds that the propagator reads 
\begin{equation}
\langle \xi^A(\sigma) \xi^B(\sigma^\prime) \rangle = H^{AB} \Delta(\sigma-\sigma^\prime) + \eta^{AB} \bar{\Delta}(\sigma-\sigma^\prime)
\end{equation}
where 
\begin{eqnarray}
\Delta(\sigma-\sigma^\prime) &=& {1 \over 2} \big( \Delta_+(\sigma-\sigma^\prime) + \Delta_-(\sigma-\sigma^\prime) \big) = - {1 \over 4 \pi} \ln (\sigma - \sigma^\prime)^2 \ , \nonumber\\ 
\bar{\Delta}(\sigma-\sigma^\prime) &=& {1 \over 2} \big( \Delta_+(\sigma-\sigma^\prime) - \Delta_-(\sigma-\sigma^\prime) \big) = - {1 \over 2 \pi} \mathrm{arctanh} {\sigma^1 - \sigma^{\prime 1} \over \sigma^0 - \sigma^{\prime 0}} \ ,
\end{eqnarray}
are the even and odd combinations of the chiral propagators \cite{Tseytlin:1990va}, with $\Delta(\sigma-\sigma^\prime)$ being the standard boson propagator. Given these relations, the single contraction term (\ref{singlecont}) is
\begin{equation}
\langle\langle \xi^A \xi^B \rangle\rangle = H^{AB} \Delta(0) + \eta^{AB} \bar{\Delta}(0)\ ,
\label{singlecont1}
\end{equation}
while the double contraction terms (\ref{doublecont}) are found by straightforward application of Wick's theorem (see the appendix of \cite{Berman:2007xn} for details) and read
\begin{eqnarray}
\langle\langle \xi^A \partial_1 \xi^B \xi^C \partial_1 \xi^D \rangle\rangle \!\!\!&\sim&\!\!\! (H^{A[C} H^{D]B} - \eta^{A[C} \eta^{D]B}) \Delta(0) \ ,\nonumber \\
\langle\langle \xi^A \partial_0 \xi^B \xi^C \partial_0 \xi^D \rangle\rangle \!\!\!&\sim&\!\!\! - ( H^{A[C} H^{D]B} + 3 \eta^{A[C} \eta^{D]B} ) \Delta(0) - 2 (  H^{A[C} \eta^{D]B} +  \eta^{A[C} H^{D]B}) \bar{\Delta}(0) \ ,\nonumber \\
\langle\langle \xi^A \partial_1 \xi^B \xi^C \partial_0 \xi^D \rangle\rangle \!\!\!&\sim&\!\!\! - ( H^{A[C} \eta^{D]B} +  \eta^{A[C} H^{D]B}) \Delta(0) - 2 \eta^{A[C} \eta^{D]B} \bar{\Delta}(0) \ ,
\label{doublecont1}
\end{eqnarray}
where $\sim$ represents equality up to finite terms that involve neither $\Delta$ nor $\bar{\Delta}$. 

As the above contractions involve propagators of fields evaluated at the same point, one needs a regularization prescription in order to handle the resulting pathologies. Namely, as $\sigma^\prime \to \sigma$, $\Delta$ diverges while $\bar{\Delta}$ becomes ambiguous because the limit can be taken by keeping ${\sigma^1 - \sigma^{\prime 1} \over \sigma^0 - \sigma^{\prime 0}}$ equal to an arbitrary constant. Regularizing $\Delta$ by sending $\ln (\sigma - \sigma^\prime)^2  \to \ln \left( (\sigma - \sigma^\prime)^2 + \mu^2 \right)$ and specifying ${\sigma^1 - \sigma^{\prime 1} \over \sigma^0 - \sigma^{\prime 0}} = \tanh\delta$, we obtain our limiting expressions
\begin{equation}
\Delta(0) \to - {1 \over 2 \pi} \ln \mu \ , \qquad \bar{\Delta}(0) \to -{{\rm 1} \over 2\pi} \delta\ .
\end{equation}
Therefore, writing the effective action coefficients $\Gamma^{\mu\nu}_{IJ}$ in (\ref{eff2terms}) as
\begin{equation}
\Gamma^{\mu\nu}_{IJ} = W^{\mu\nu}_{IJ} \Delta(0) + L^{\mu\nu}_{IJ} \bar{\Delta}(0)\ ,
\label{weyllorentz}
\end{equation}
we see that the first term introduces a dependence on the scale $\mu$ indicating breakdown of scale invariance (with the coefficient $W^{\mu\nu}_{IJ}$ being identified with the Weyl anomaly), while the second term introduces a dependence on the ``boost parameter'' $\delta$ signifying breakdown of Lorentz invariance (with the coefficient $L^{\mu\nu}_{IJ}$ being identified with the so-called global Lorentz anomaly). If these models are to describe consistent string backgrounds, we should require that the corresponding parts of the effective action,
\begin{eqnarray}
W &=& {1 \over 2} \int d^2 \sigma \left( W^{00}_{IJ} \partial_0 {\mathbb Y}^I \partial_0 {\mathbb Y}^J + W^{01}_{IJ} \partial_0 {\mathbb Y}^I \partial_1 {\mathbb Y}^J + W^{11}_{IJ} \partial_1 {\mathbb Y}^I \partial_1 {\mathbb Y}^J \right)\ , \nonumber\\ 
L &=& {1 \over 2} \int d^2 \sigma \left(L^{00}_{IJ} \partial_0 {\mathbb Y}^I \partial_0 {\mathbb Y}^J + L^{01}_{IJ} \partial_0 {\mathbb Y}^I \partial_1 {\mathbb Y}^J + L^{11}_{IJ} \partial_1 {\mathbb Y}^I \partial_1 {\mathbb Y}^J \right)\ , 
\label{fullweylandlorentz}
\end{eqnarray}
both vanish on shell, either identically or by imposing appropriate restrictions on the background fields in the form of equations of motion or constraints. 
 
\subsection{Final expressions}

To conclude this section, we will use the above results to write explicit expressions for the Weyl and Lorentz anomalies and we will present the quantities ${\cal S}$, ${\cal Q}$ and ${\cal P}$. Substituting the contractions (\ref{singlecont1}) and (\ref{doublecont1})
in (\ref{eff2terms}) and using the decomposition (\ref{weyllorentz}), we find that the Weyl anomaly is given by
\begin{eqnarray}
W^{00}_{IJ} &=& {1 \over 4} (H^{A[C} H^{D]B} - \eta^{A[C} \eta^{D]B}) {\cal Q}^{0}_{AB,I} {\cal Q}^{0}_{CD,J}\ ,\nonumber\\
W^{01}_{IJ} &=& H^{AB} {\cal S}^{01}_{AB,IJ} + {1 \over 2} (H^{A[C} H^{D]B} - \eta^{A[C} \eta^{D]B}) {\cal Q}^{0}_{AB,I} {\cal Q}^{1}_{CD,J} \nonumber\\
&-& {1 \over 4} ( H^{A[C} \eta^{D]B} +  \eta^{A[C} H^{D]B}) \left( {\cal Q}^{0}_{AB,I} {\cal P}^{1}_{CD,J} + {\cal P}^{1}_{AB,I} {\cal Q}^{0}_{CD,J} \right)\ , \nonumber\\
W^{11}_{IJ} &=& H^{AB} {\cal S}^{11}_{AB,IJ} + {1 \over 4} (H^{A[C} H^{D]B} - \eta^{A[C} \eta^{D]B}) {\cal Q}^{1}_{AB,I} {\cal Q}^{1}_{CD,J} \nonumber\\
&-& {1 \over 4} ( H^{A[C} \eta^{D]B} +  \eta^{A[C} H^{D]B}) \left( {\cal Q}^{1}_{AB,I} {\cal P}^{1}_{CD,J} + {\cal P}^{1}_{AB,I} {\cal Q}^{1}_{CD,J} \right) \nonumber\\
&-& {1 \over 4} ( H^{A[C} H^{D]B} + 3 \eta^{A[C} \eta^{D]B} ) {\cal P}^{1}_{AB,I} {\cal P}^{1}_{CD,J} \ ,
\label{weylfinal}
\end{eqnarray}
while for the Lorentz anomaly we find $L^{00}_{IJ}=0$ and
\begin{eqnarray}
L^{01}_{IJ} &=& \eta^{AB} {\cal S}^{01}_{AB,IJ} -  {1 \over 2} \eta^{A[C} \eta^{D]B} \left( {\cal Q}^{0}_{AB,I} {\cal P}^{1}_{CD,J} + {\cal P}^{1}_{AB,I} {\cal Q}^{0}_{CD,J} \right)\ ,\nonumber\\
L^{11}_{IJ} &=& \eta^{AB} {\cal S}^{11}_{AB,IJ} - {1 \over 2} (  H^{A[C} \eta^{D]B} +  \eta^{A[C} H^{D]B} ) {\cal P}^{1}_{AB,I} {\cal P}^{1}_{CD,J} \nonumber\\
&-& {1 \over 2} \eta^{A[C} \eta^{D]B} \left( {\cal Q}^{1}_{AB,I} {\cal P}^{1}_{CD,J} + {\cal P}^{1}_{AB,I} {\cal Q}^{1}_{CD,J} \right)\ .
\label{lorentzfinal}
\end{eqnarray}

Eqs.~(\ref{weylfinal}) and (\ref{lorentzfinal}) are completely general and valid for any expansion of the action (\ref{icb}). To apply them
to the case at hand, we need explicit expressions for the quantities ${\cal S}$, ${\cal Q}$ and ${\cal P}$ introduced in eqs.~(\ref{interaction})
and (\ref{sqp}). These are read off from the expression that results when we write the second-order action (\ref{secordetacov}) or (\ref{secordHcov}) in terms of the tangent-space fields $\xi^A$, expand the worldsheet covariant derivatives and ignore the kinetic terms present in (\ref{kinetic}). The results, in the $\eta$-covariant and the $H$-covariant formulations, are presented below.
\vskip .2cm
\noindent $\bullet$ {\bf \boldmath{$\eta$}-covariant formulation.} For this case, the quantities in (\ref{eff2terms}) are given by
\begin{eqnarray}
{\cal S}^{11}_{AB,IJ} &=& - {1 \over 2} D_A D_B H_{IJ}- {1 \over 2} ( H_{IK}  
R^{K}_{\phantom{K}ABJ} + H_{JK} R^{K}_{\phantom{K}ABI} ) - H_{CD} \Omega_{I\phantom{C}A}^{\phantom{I} C} \Omega_{J\phantom{D}B}^{\phantom{J} D} \nonumber\\
&-& D_A H_{CI} \Omega_{J\phantom{C}B}^{\phantom{J} C} - D_A H_{CJ} \Omega_{I\phantom{C}B}^{\phantom{I} C}\ ,\nonumber\\
{\cal S}^{01}_{AB,IJ} &=& R_{IABJ} + {1 \over 2} D_B G_{AIJ} + \eta_{CD} \Omega_{I\phantom{C}A}^{\phantom{I} C} \Omega_{J\phantom{D}B}^{\phantom{J} D}
+\frac{1}{2} (G_{IDA} \Omega_{J\phantom{D}B}^{\phantom{J}D}-
G_{JDA} \Omega_{I\phantom{D}B}^{\phantom{J}D})
\ ,\nonumber\\
{\cal Q}^{1}_{AB,I} &=& -2 \Omega_{I\phantom{C}A}^{\phantom{I} C} H_{CB} - 2 D_A H_{BI}\ ,\nonumber\\
{\cal Q}^{0}_{AB,I} &=& - {1 \over 2} G_{IAB} + \Omega_{I\phantom{C}A}^{\phantom{I} C} \eta_{CB} \ ,\nonumber\\
{\cal P}^{1}_{AB,I} &=& {1 \over 2} G_{IAB} + \Omega_{I\phantom{C}A}^{\phantom{I} C} \eta_{CB}\ ,
\label{sqpeta}
\end{eqnarray}
where $\Omega_{I\phantom{A}B}^{\phantom{I} A}$ is the spin connection compatible with $\eta_{IJ}$.
 \vskip .2cm
\noindent $\bullet$ {\bf \boldmath{$H$}-covariant formulation.} The quantities in (\ref{eff2terms}) read
\begin{eqnarray}
{\cal S}^{11}_{AB,IJ} &=& - R_{IABJ} - H_{CD} \Omega_{I\phantom{C}A}^{\phantom{I} C} \Omega_{J\phantom{D}B}^{\phantom{J} D}\ ,\nonumber\\
{\cal S}^{01}_{AB,IJ} &=& {1 \over 2} D_B G_{AIJ} + {1 \over 2} D_A D_B \eta_{IJ} + {1 \over 2} (\eta_{IK} R^{K}_{\phantom{K}ABJ} + \eta_{JK} R^{K}_{\phantom{K}ABI} ) \nonumber\\ 
&+& \left( D_A \eta_{CI} - {1 \over 2} G_{IAC} \right) \Omega_{J \phantom{C} B}^{\phantom{J} C} + \left( D_A \eta_{CJ} + {1 \over 2} G_{JAC} \right) \Omega_{I \phantom{C} B}^{\phantom{I} C} +  \eta_{CD} \Omega_{I\phantom{C}A}^{\phantom{I} C} \Omega_{J\phantom{D}B}^{\phantom{J} D}\ ,\nonumber\\
{\cal Q}^{1}_{AB,I} &=& -2 \Omega_{I\phantom{C}A}^{\phantom{I} C} H_{CB}\ ,\nonumber\\
{\cal Q}^{0}_{AB,I} &=& D_A \eta_{BI} - {1 \over 2} G_{IAB} + \Omega_{I\phantom{C}A}^{\phantom{I} C} \eta_{CB}\ ,\nonumber\\
{\cal P}^{1}_{AB,I} &=& D_A \eta_{BI} + {1 \over 2} G_{IAB} + \Omega_{I\phantom{C}A}^{\phantom{I} C} \eta_{CB}\ ,
\label{sqpH}
\end{eqnarray}
where now $\Omega_{I\phantom{A}B}^{\phantom{I} A}$ is the spin connection compatible with $H_{IJ}$.
\vskip .2cm

Finally, let us remark that, since the fluctuations $\xi^A$ have been integrated out, the formulas given above are obviously independent of the precise form of $\eta_{AB}$ and $H_{AB}$ as long as the tangent space version of the Lorentz invariance condition (\ref{lorentz1}) is obeyed; the particular choice of the chiral basis (\ref{vb2}) was made only for the purpose of simplifying the form of the kinetic Lagrangian and of
the propagators deriving from it. In what follows, we will change our basis and we will employ the standard convention for the choice of $\eta_{AB}$, given by eq.~(\ref{etaspec}) below.

\section{Conformal models on twisted doubled tori}
\label{conformaltdt}

We now restrict ourselves to a specific class of backgrounds, namely the twisted doubled tori (TDT) which were introduced in \cite{Dall'Agata:2007sr} as candidate internal (doubled) geometries underlying a broad class of ${\cal N}=4$ gauged supergravity theories. We begin by reviewing the motivation for the introduction of these backgrounds and then, following \cite{Dall'Agata:2008qz}, we formulate the interacting chiral boson theory on these backgrounds and explain why it is Lorentz invariant. Next, we apply the results of the previous section for the calculation of the Weyl and Lorentz anomalies to these models, in order to determine which of them correspond to consistent string vacua. We demonstrate that the condition for the vanishing of the Weyl anomaly is equivalent to the minimization condition for the scalar potential of the associated gauged supergravity. The interesting fact that all compact gaugings lead to TDT sigma-models which are conformal is explained by demonstrating that these models are actually chiral WZW models. We also investigate the possible existence of conformally invariant models in a series of simple examples where certain
combinations of fluxes are turned on.

\subsection{Twisted doubled tori and \boldmath{${\cal N}=4$} gauged supergravities }

As stated in the introduction, the idea of doubling the coordinates is quite natural once one wants to treat momentum and winding modes on an equal footing. For a compactification on a $d$-dimensional torus parametrized by coordinates $y^i$, this doubling amounts to considering a $2d$-dimensional {\em doubled torus} parametrized by $\{y^i, y^{\tilde i}\}$ \cite{Witten:1988zd, Duff:1989tf, Tseytlin:1990nb, Tseytlin:1990va, Maharana:1992my, Hull:2004in, Hull:2006va, Hull:2007zu}. An elementary example is provided by the usual reduction of the NS-NS sector of string theories on $\mathbb{T}^d$. The moduli fields originating from the metric and the B-field can be assembled into the generalized metric
\begin{equation}
M_{AB} = \left( 
\begin{array}{cc}
	G_{ab} - B_{ac}G^{cd}B_{db} & B_{ac}G^{cb}\\
	-G^{ac}B_{cb} & G^{ab} 
\end{array}
\right) \ ,
\label{genmetr} 
\end{equation}
and the reduction yields an ungauged supergravity theory with no potential for the moduli. This effective supergravity theory could be thought of as a reduction of an appropriate double field theory on the doubled torus, whose isometry group results in the full abelian 
gauge symmetry group ${\rm U}(1)^{2d}$.

To arrive at a mechanism for obtaining gauged supergravity theories from the doubled formalism, it is instructive to first consider the toroidal reduction of a purely gravitational theory. The standard way to obtain a moduli potential as well as non-abelian gauge interactions is to perform a Scherk--Schwarz reduction \cite{Scherk:1979zr}, {\it i.e.}~turn the $d$-dimensional torus into a local group manifold by twisting and then reduce the higher-dimensional theory in such a way that only the left-translation isometries remain as gauge symmetries in the effective theory. The resulting potential has the schematic form
\begin{equation}
V_{SS} \sim  2 \tau_{da}^{\phantom{da}c} \tau_{cb}^{\phantom{cb}d} G^{ab} + \tau_{ce}^{\phantom{ce}a} \tau_{df}^{\phantom{df}b} G_{ab} G^{cd} G^{ef}\ ,
\label{sspot} 
\end{equation}
where $G_{ab}$ are the moduli coming from the metric and the geometric flux $\tau_{ab}^{\phantom{ab}c}$ corresponds to the structure constants of the group manifold defined by the twisting.

The above construction may already suggest the possibility that twisting the doubled torus and reducing an appropriate double field theory on the resulting local group manifold should yield gauged supergravity theories with non-abelian gauge groups and potentials for the moduli. The crucial evidence that motivates  this idea in more concrete terms \cite{Dall'Agata:2007sr} is that the potential of such theories has a form completely analogous to the Scherk--Schwarz potential (\ref{sspot}), namely
\begin{equation}
V \sim 2 \, {\cal T}_{DA}^{\phantom{DA}C} {\cal T}_{CB}^{\phantom{CB}D} M^{AB} + {\cal T}_{CE}^{\phantom{CE}A} {\cal T}_{DF}^{\phantom{DF}B}  M_{AB} M^{CD} M^{EF}\ .
\label{ssgenpot}
\end{equation}
Here, $M^{AB}$ are the moduli fields while the embedding tensor ${\cal T}_{AB}^{\phantom{AB}C}$ parametrizes the gauging and, for the theories considered in this paper,\footnote{The precise relationship between the embedding tensor and the structure constants of the gauge algebra is more complicated in general \cite{deWit:2005ub}.} actually corresponds to the structure constants of the gauge algebra, {\it i.e.}
\begin{equation}
\left[{\mathbb X}_A,{\mathbb X}_B\right] = {\cal T}_{AB}{}^C {\mathbb X}_C\ .
\label{gaugealgebra}
\end{equation}
The gaugings described by the embedding tensor of these theories are more general than those that can be accounted for by the conventional Scherk--Schwarz geometric fluxes and the $p$-form fluxes of higher-dimensional theories. Therefore, one might expect that the twisted doubled torus construction might provide us with a lift of the supergravity theories under consideration, for any of the allowed gaugings, to higher dimensions. The higher-dimensional origin of these gaugings should involve all types of fluxes that can be described by ${\cal T}_{AB}^{\phantom{AB}C}$, namely the physical ones ({\it e.g.}~NS-NS three-form flux), the geometric ones ({\it i.e.}~twisted tori) as well as the non-geometric ones. 

The effective supergravities that will be considered in this paper are ${\cal N}=4$ gauged supergravities in four spacetime dimensions \cite{Gates:1982ct, de Roo:1984gd, deRoo:1985jh, deRoo:1986yw, Bergshoeff:1985ms}.
Such theories were constructed
in full generality in \cite{Schon:2006kz} using the embedding tensor formalism. A key point of these theories is the existence of a global ${\rm SL}(2,\mathbb{R}) \times {\rm SO}(6,d)$ duality symmetry, where  $d$ is  the number of vector multiplets. The first factor of this group contains the axionic shift as well as the electric-magnetic duality transformations while the second factor is the T-duality group.
 
The ${\cal N}=4$ gaugings are parametrized by two sets of tensors $f_{\alpha ABC}$ and $\xi_{\alpha A}$, where $\alpha$ labels doublets of the ${\rm SL}(2,\mathbb{R})$ duality group while $A,B,\ldots$ are in the fundamental of ${\rm SO}(6,d)$ and are fully antisymmetrized. The most natural way, {\it i.e.}~without invoking orientifolds or similar constructions, to embed these theories in higher dimensions is through heterotic string theory reduced on $\mathbb{T}^6$. Then we obtain twelve gauge fields, six originating from the reduction of the metric and six from the reduction of the B-field (we ignore the gauge fields coming from the ${\rm SO}(32)$ or $E_8 \times E_8$ sector) and hence we have $d=6$. Still, keeping in mind tori of generic dimension $d$, we will consider the more general theories with ${\rm O}(d,d)$ T-duality group.

The gaugings that are the focal point of this paper are the so-called electric ones where only one set of the $f_{\alpha ABC}$, for instance those with $\alpha=1$, is taken to be nonzero. Then, consistency of the gauging requires that $f_{1 ABC}$ satisfy Jacobi identities and that the ${\rm O}(d,d)$ metric $\eta_{AB}$ be an invariant metric of the gauge algebra (see \cite{Schon:2006kz, Derendinger:2007xp} for more information on gaugings of this type). In this case, the gauge algebra is (\ref{gaugealgebra}) with ${\cal T}_{AB}^{\phantom{AB}D} \eta _{CD} = {\cal T}_{ABC} \equiv f_{1ABC}$.

The higher-dimensional origin of such gaugings and the possible interpretation of the embedding tensor in terms of fluxes has been discussed in \cite{Schon:2006kz, Derendinger:2007xp, Dall'Agata:2007sr, ReidEdwards:2008rd, Aldazabal:2008zza, Dall'Agata:2009gv}. Based on a comparison of the potential and  the gauge algebras, it was suggested in \cite{Derendinger:2007xp, Dall'Agata:2007sr} that for electric gaugings, $f_{1 ABC}$ comprises of geometric flux, NS-NS flux as well as their T-dual $Q$- and $R$-flux \cite{Shelton:2005cf, Aldazabal:2006up,Shelton:2006fd}. This set of fluxes is closed under T-duality transformations and the main proposal of \cite{ Dall'Agata:2007sr} was that they can all be considered as geometric flux on the twisted doubled torus.

It will be instructive and useful for later applications to show explicitly how the embedding tensor encodes the various types of fluxes. For this purpose we separate the generators of the gauge algebra (\ref{gaugealgebra}) as ${\mathbb X}_A=\{Z_a,X^a\}$, with $Z_a$ and $X^a$ corresponding to the gauge fields coming from the metric and the B-field respectively, and decompose the embedding tensor accordingly. Then, (\ref{gaugealgebra}) can be rewritten as
\begin{eqnarray}
\text{[}Z_a , Z_b] &=& \tau_{ab}^{\phantom{ab}c} Z_c + H_{abc} X^c\ , \nonumber \\
\text{[} X^a , Z_b ] &=& \tau_{bc}^{\phantom{bc}a} X^c - Q_b^{\phantom{b}ac} Z_c\ , \label{thqr} \\
\text{[}X^a , X^b] &=& Q_c^{\phantom{c}ab} X^c + R^{abc} Z_c\ .\nonumber
\end{eqnarray}

Now, the ${\cal N}=4$ gauged supergravity potential for this class of gaugings reads\footnote{We have dropped the dilaton dependence, since it is not taken into account in the worldsheet theory.} \cite{Schon:2006kz, Derendinger:2007xp}
 \begin{equation}
V(M) = {1 \over 2} \left(\frac{1}{3} M^{AA'} M^{BB'} M^{CC'} + \left( \frac{2}{3} \eta^{AA'} - M^{AA'} \right) \eta^{BB'} \eta^{CC'} \right) {\cal T}_{ABC} {\cal T}_{A'B'C'}\ ,
\label{gaugedsugpo}
\end{equation}
where the symmetric matrix $M^{AB}$ parametrizes ${{\rm O}(d,d) \over {\rm O}(d) \times {\rm O}(d)}$ and hence satisfies
\begin{equation}
\eta_{AB} = M_{AC} \eta^{CD} M_{DB}\ ,
\label{lor}
\end{equation}
with $M_{AB}$ being the inverse of $M^{AB}$. Comparing this potential, using the decomposition (\ref{thqr}) and the parametrization (\ref{genmetr}), with those obtained from  Scherk-Schwarz reductions with fluxes \cite{Scherk:1979zr, Kaloper:1999yr} as well as with their non-geometric generalizations  \cite{Shelton:2005cf, Aldazabal:2006up, Shelton:2006fd}, leads us to identify $H_{abc}$ with the standard NS-NS flux, $\tau_{ab}^{\phantom{ab}c}$ with the geometric flux, $Q_c^{\phantom{c}ab}$ with the locally geometric flux and $R^{abc}$ with the so-called non-geometric flux. 

The program of explicitly reducing higher-dimensional theories on twisted doubled tori in the manner outlined above cannot be explicitly performed, since at this point the double theories to be reduced are yet unknown. The models presented in \cite{Hull:2009mi} might be an appropriate starting point but one still has to await for their full nonlinear completion in order to be able to compare with the generic gauged supergravity theory. Given this state of affairs, one is compelled to resort to the complementary worldsheet approach.

\subsection{Lorentz invariant sigma-models on twisted doubled tori}
 
In the absence of a candidate spacetime theory that should be reduced on a TDT to yield the gauged supergravity theories under consideration, one can consider the TDT as target spaces of appropriate two-dimensional theories and hope that they can be elevated to the status of an actual string background. Since these are doubled geometries, the natural model to consider is the interacting chiral boson model (\ref{icb}) presented in Section 2, as it naturally accommodates worldsheet theories where momentum and winding modes are treated in a democratic, {\it i.e.}~${\rm O}(d,d)$ covariant, fashion. 

A first step in this direction was taken  in \cite{Dall'Agata:2008qz}, where it was shown that with appropriate choices of the metrics $H_{IJ}, \eta_{IJ}$ and the generalized flux $G_{IJK}$, specified by the gauging parameters  ${\cal T}_{AB}^{\phantom{AB}C}$ and naturally related to the geometry of the TDT, the corresponding chiral boson models have (on-shell) two-dimensional Lorentz invariance. The starting point is the gauge algebra ${\mathfrak g}$ 
with generators ${\mathbb X}_A$ satisfying (\ref{gaugealgebra}). As we have emphasized, the structure constants $ {\cal T}_{AB}^{\phantom{AB}C}$ are identified with the embedding tensor of gauged supergravity which, for the case under consideration, 
satisfies the usual Jacobi identities and fully determines the gauging. 

We can now select a group representative $g({\mathbb Y}) = {\rm exp}({\mathbb Y}^I  {\mathbb X}_I)$, where $ {\mathbb X}_I$ should be taken in a faithful representation\footnote{Notice that
the generic gauge algebra  ${\mathfrak g}$ may contain a non-trivial abelian ideal ${\cal I}$ whose generators in the adjoint representation are embedded trivially in the ${\mathfrak o}(d,d)$ algebra of the duality group. For instance, in the case of no gauging at all, {\it i.e.}~${\cal G} = {\rm U(1)}^{2d}$ with $ {\cal T}_{AB}{}^C =0$, the adjoint generators are all zero and in order to construct a non-trivial group element we obviously need a faithful representation of the gauge algebra. Such a faithful representation exists for any ${\mathfrak g}$, as
it is a general result that all finite-dimensional Lie algebras admit
faithful finite-dimensional representations.} 
of ${\mathfrak g}$, and construct the left-invariant vielbein $E^A$ as $E^A {\mathbb X}_A = g^{-1} dg$ which satisfies the Maurer-Cartan structure equations
\begin{equation}
d E^A = -{1 \over 2} {\cal T}_{BC}^{\phantom{BC} A} E^B \wedge E^C\ .
\label{structureeq}
\end{equation}
The doubled vielbein can be written as
\begin{equation}
E^A = E^A_I d {\mathbb Y}^I\ ,
\end{equation}
with the matrix $E^A_I$ playing the role of the Scherk--Schwarz twist matrix in the doubled formalism. For that reason, the corresponding local group manifold ${\cal G} / \Gamma$, parametrized by the coordinates ${\mathbb Y}^I$ is called {\em twisted doubled torus}. Notice that in order to obtain a compact space we might need to compactify with respect to the action of a discrete cocompact subgroup $\Gamma$ of the left-translation isometries.
This quotient leaves the
above vielbein invariant but imposes global identifications.

Using the vielbein given above, we construct two different metrics, namely
\begin{equation}
\eta_{IJ} = \eta_{AB} E^A_I E^B_J\ ;\qquad \eta_{AB} = \left(
\begin{array}{cc}
	0 & \mathbbm{1}_d \\
	\mathbbm{1}_d & 0 
\end{array}
 \right)\ ,
\label{etaspec}
\end{equation}
and
\begin{equation}
H_{IJ} = H_{AB} E^A_I E^B_J\ ,
\label{hspec}
\end{equation}
with the tangent space metric $H_{AB}$ subject to the requirement that it belongs to the coset ${{\rm O}(d,d) \over {\rm O}(d)\times {\rm O}(d)}$, {\it i.e.}~it  satisfies
\begin{equation}
\eta_{AB} = H_{AC} \eta^{CD} H_{DB}\ .
\label{tanh}
\end{equation}
The requirement that the adjoint representation of the gauge algebra  ${\mathfrak g}$ is embedded in the fundamental representation of the duality algebra ${\mathfrak o}(d,d)$ \cite{deWit:2005ub} implies that the structure constants obey
\begin{equation}
{\cal T}_{AB}^{\phantom{AB} D} \eta_{DC} = - {\cal T}_{AC}^{\phantom{AC} D} \eta_{DB}\ .
\label{structureconstantsodd}
\end{equation}
Taking $\eta_{AB}$ as the tangent space metric, the compatible spin connection satisfies
\begin{equation}
\Omega^C_{\phantom{C}A} \eta_{CB} + \Omega^C_{\phantom{C}B} \eta_{CA} = 0\ .
\end{equation}
From (\ref{structureeq}), the general solution for the spin connection is
\begin{equation}
\Omega_{AB} = \frac{1}{2} ({\cal T}_{ABC} + {\cal T}_{ACB}- {\cal T}_{BCA}) E^C,
\end{equation}
where ${\cal T}_{ABC} =  {\cal T}_{AB}^{\phantom{AB} D} \eta_{DC}$. The antisymmetry property (\ref{structureconstantsodd}) allows us to rewrite this as
\begin{equation}
 \Omega^A_{\phantom{A}B}= - \frac{1}{2} {\cal T}_{BC}^{\phantom{BC}A} E^C \quad \textrm{or} \quad   \Omega_{AB}^{\phantom{AB}C}  = -{1 \over 2} {\cal T}_{AB}^{\phantom{AB} C} \ ,
\label{spinconnectionTDT} 
\end{equation}
where $\Omega^A_{\phantom{A}B} = \Omega^{\phantom{I}A}_{I\phantom{A}B} d\mathbb{Y}^I = \Omega^{\phantom{C}A}_{C\phantom{A}B}
\mathbb{E}^C$. The curvature of the TDT reads
\begin{equation}
R^A_{\phantom{A}B} = \frac{1}{4} {\cal T}_{CB}^{\phantom{CB}E}
{\cal T}_{DE}^{\phantom{DE}A}
 E^C \wedge E^D\quad \textrm{or} \quad R^A_{\phantom{A} BCD} = \frac{1}{4}  {\cal T}_{EB}^{\phantom{EB}A}
 {\cal T}_{DC}^{\phantom{DC}E}\ ,
\label{riemannTDT}
\end{equation}
where $R^A_{\phantom{A}B}$ is the Ricci 2-form and $R^A_{\phantom{A} BCD}$ is the Riemann tensor. We should mention that in the above manipulations we have used repeatedly the Jacobi identity for ${\cal T}_{AB}^{\phantom{AB}C}$.

The main result of ref.~\cite{Dall'Agata:2008qz} was that if we consider the model (\ref{icb}) with background fields specified by eqs.~(\ref{etaspec}) and (\ref{hspec}) and we further turn on a generalized flux
\begin{equation}
G_{ABC} = {\cal T}_{ABC}\ ,
\label{htplussign}
\end{equation} 
then the Lorentz invariance conditions (\ref{lorentz1}) and (\ref{lorentz2}) are satisfied. More specifically, (\ref{lorentz1}) is true by construction while the classical equations of motion can be rewritten in the form $( D_I V_J - {1 \over 2} {\cal T}_{IJ}^{\phantom{IJ}K} V_K) \partial_1 {\mathbb Y}^I=0$, which leads to (\ref{lorentz2}) upon contraction with $\eta^{JL}V_L$. 

Moreover, for compact gaugings, {\it i.e.}~for structure constants that in addition to (\ref{structureconstantsodd}) obey
\begin{equation}
{\cal T}_{AB}{}^D H_{DC} = - {\cal T}_{AC}{}^D H_{DB}\ ,
\label{compgau}
\end{equation}
it was shown in \cite{Dall'Agata:2008qz} that classical Lorentz invariance also holds for the alternative choice of generalized flux
\begin{equation}
G_{ABC} = - {\cal T}_{ABC}\ .
\label{htminussign}
\end{equation} 
In this case the equations of motion reduce to the first-order form $V_I=0$ which trivially satisfies eq.~(\ref{lorentz2}). In both cases, the crucial characteristic that makes the TDT work is the fact that it is a local group manifold and therefore the spin connection compatible with the invariant metric is fully antisymmetric. Therefore, the three-form flux given by (\ref{htplussign}) or (\ref{htminussign}) can act as a torsion piece that parallelizes this connection and brings the equations of motion to a simplified form that entails the Lorentz invariance condition. 

\subsection{Weyl and Lorentz anomalies}

Having established the existence of a class of backgrounds which are associated to supergravity gaugings and lead to Lorentz invariant interacting chiral boson theories, it is crucial to examine whether they correspond to consistent string vacua. This amounts to computing the Weyl anomaly, whose non-vanishing signals the breakdown of conformal invariance in the quantum theory, as well and the Lorentz anomaly arising due to the chirality of the bosons in the sigma-model. For the class of models under consideration, this should yield conditions on the structure constants ${\cal T}_{AB}^{\phantom{AB}C} $ that distinguish those TDT that are actual string theory backgrounds.

For the computation we will employ the general machinery developed in the previous section, which culminated in the expressions (\ref{weylfinal}) and (\ref{lorentzfinal}) for the Weyl and Lorentz anomalies respectively. Although both the $\eta$-covariant and $H$-covariant formulations of the background expansion might in principle be used, the facts that $\eta_{AB}$ is the invariant tangent space metric and that the structure constants obey 
(\ref{structureconstantsodd}) indicate that it is most appropriate to use the $\eta$-covariant expansion, {\it i.e.}~eq.~(\ref{sqpeta}).

Based on standard results on the ultraviolet properties of standard 
sigma-models \cite{Curtright:1984dz, Braaten:1985is, Mukhi:1985vy}
on group manifolds, we are led to expect that the choices of generalized three-form flux given in (\ref{htplussign}) or (\ref{htminussign}), which parallelize the connection and lead to Lorentz-invariant theories, will lead also to simple expressions for the Weyl anomaly. This is also evident by inspection of eq.~(\ref{sqpeta}), whose various terms simplify considerably once we use eqs.~(\ref{spinconnectionTDT}) and (\ref{riemannTDT}) and impose (\ref{htplussign}) or (\ref{htminussign}). To illustrate this point in detail, we will not impose classical Lorentz invariance from the beginning, but we will consider a more general choice for the generalized flux, namely
\begin{equation}
G_{ABC} = \lambda {\cal T}_{ABC}\ .
\label{lahtplussign}
\end{equation} 

To present the relevant formulas without cluttering the notation, we will adopt the convention that structure constant indices raised and lowered with $H$ will be represented with a hat, 
e.g.~$T_{AB\hat{C}}=T_{AB}^{\phantom{AB}D} H_{DC}$, $T_{AB\hat{I}} = T_{AB}^{\phantom{AB}C}  H_{CD} E^D_I$, etc\dots 
Then, substitution of eqs.~(\ref{lahtplussign}) and (\ref{spinconnectionTDT}) in eq.~(\ref{sqpeta}) yields
\begin{eqnarray}
{\cal S}^{11}_{AB,IJ} &=& \frac{1}{2} (  {\cal T}_{\hat{I}AC} {\cal T}_{JB}^{\phantom{JB}C} + {\cal T}_{\hat{J}AC} {\cal T}_{IB}^{\phantom{JB}C} ) - \frac{1}{2} ( {\cal T}_{IA\hat{C}} {\cal T}_{JB}^{\phantom{JB}C} + {\cal T}_{JA\hat{C}} {\cal T}_{IB}^{\phantom{IB}C} )\ , \nonumber\\ 
{\cal S}^{01}_{AB,IJ} &=& - \frac{\lambda}{4} {\cal T}_{IJ}^{\phantom{IJ}C} {\cal T}_{ABC} \ , \nonumber\\
{\cal Q}^{1}_{AB,I} &=& - 2 {\cal T}_{IA\hat{B}} + {\cal T}_{\hat{I}AB}\ , \nonumber\\
{\cal Q}^{0}_{AB,I} &=& -\frac{\lambda-1}{2} {\cal T}_{IAB}\ , \nonumber\\
{\cal P}^1_{AB,I} & =&\frac{\lambda+1}{2}{\cal T}_{IAB} \ ,
\label{sqpTDT}
\end{eqnarray}
where the Jacobi identity for the structure constants ${\cal T}_{AB}^{\phantom{AB}C}$ has been repeatedly used. Plugging this expression in eqs.~(\ref{weylfinal}) and (\ref{lorentzfinal}), we find that the coefficients of the Weyl anomaly are
\begin{eqnarray}
W^{00}_{IJ} &=& \frac{(\lambda-1)^2}{16} ({\cal T}_{IAB} {\cal T}_J^{\phantom{J}AB} - {\cal T}_{IAB} {\cal T}_J^{\phantom{J}\hat{A}\hat{B}})\ , \nonumber \\
W^{01}_{IJ}&=& \frac{\lambda-1}{4} \left( {\cal T}_{IAB} {\cal T}_{\hat{J}}^{\phantom{\hat{J}}AB} - {\cal T}_{IAB} {\cal T}_{\hat{J}}^{\phantom{\hat{J}}\hat{A}\hat{B}}
+(\lambda+1) {\cal T}_{IAB} {\cal T}_J^{\phantom{J}\hat{A}B} \right)\ ,\nonumber  \\
W^{11}_{IJ} &=& {(\lambda+1)\left(8-3(\lambda+1)\right) \over 16} {\cal T}_{IAB} {\cal T}_J^{\phantom{J}AB} + {(\lambda+1)\left(8-(\lambda+1)\right) - 16 \over 16} {\cal T}_{IAB} {\cal T}_J^{\phantom{J}\hat{A}\hat{B}}
\nonumber\\
&-& \frac{1}{4} ( {\cal T}_{\hat{I}AB} {\cal T}_{\hat{J}}^{\phantom{\hat{J}}AB} - {\cal T}_{\hat{I}AB} {\cal T}_{\hat{J}}^{\phantom{\hat{J}}\hat{A}\hat{B}} ) - \frac{\lambda-1}{4} ( {\cal T}_{\hat{I}AB} {\cal T}_J^{\phantom{J}\hat{A}B} + {\cal T}_{\hat{J}AB} {\cal T}_I^{\phantom{I}\hat{A}B})\ ,
\label{weyltdt}
\end{eqnarray}
while those of the Lorentz anomaly read
\begin{eqnarray}
L^{01}_{IJ} &=& \frac{\lambda^2-1}{4} {\cal T}_{IAB} {\cal T}_J^{\phantom{J}AB}\ , \nonumber\\
L^{11}_{IJ} &=& - \frac{(\lambda-1)^2}{4} {\cal T}_{I\hat{A}B} {\cal T}_{J}^{\phantom{J}AB} - \frac{\lambda-1}{4} ( {\cal T}_{\hat{I}AB} {\cal T}_J^{\phantom{J}AB} + {\cal T}_{\hat{J}AB} {\cal T}_I^{\phantom{I}AB} )\ .
\label{lorentztdt}
\end{eqnarray}

For generic $\lambda$, the above contributions are non-vanishing and the full expressions (\ref{fullweylandlorentz}) for the Weyl and Lorentz anomalies cannot be brought to any meaningful form by manipulating them using the sigma-model equations of motion. However, 
for $\lambda=\pm 1$, the equations simplify considerably, as anticipated. We examine the two cases in turn.

\vskip .2cm
\noindent $\bullet$ {\bf $\lambda=1$.} For this case, where classical Lorentz invariance is guaranteed for both compact and non-compact gaugings, it is immediately seen that all coefficients of the Lorentz anomaly vanish, {\it i.e.}~that quantum Lorentz invariance follows once it is established at the classical level. As for the Weyl anomaly, the first two coefficients in (\ref{weyltdt}) vanish, while the third can be rearranged as
\begin{equation}
W^{11}_{IJ} = \frac{1}{4} \left( {\cal T}_{IAB} {\cal T}_J^{\phantom{J}AB} - {\cal T}_{IAB} {\cal T}_J^{\phantom{J}\hat{A}\hat{B}} - {\cal T}_{\hat{I}AB} {\cal T}_{\hat{J}}^{\phantom{\hat{J}}AB} + {\cal T}_{\hat{I}AB} {\cal T}_{\hat{J}}^{\phantom{\hat{J}}\hat{A}\hat{B}} \right)\ ,
\label{w11TDT0}
\end{equation}
or, more explicitly, in tangent space components
\begin{equation}
W^{11}_{AB} = \frac{1}{4}  ( \eta_{AE} \eta_{BE'} - H_{AE} H_{BE'} ) ( \eta^{CC'} \eta^{DD'} - H^{CC'} H^{DD'} ) {\cal T}_{CD}^{\phantom{CD}E} {\cal T}_{C'D'}^{\phantom{C'D'}E'}\ .
\label{w11TDT}
\end{equation}
  For a non-compact gauging, the vanishing of this quantity yields a condition on the structure constants of the TDT that is sufficient for obtaining a (one-loop) conformally invariant sigma-model. For a compact gauging, we can furthermore use the condition (\ref{compgau}) to arrive at the relation
\begin{equation}
 \eta^{CC'} \eta^{DD'} {\cal T}_{ACD} {\cal T}_{BC'D'} = 
H^{CC'} H^{DD'} {\cal T}_{ACD} {\cal T}_{BC'D'} \ , 
\label{HHetaeta}
\end{equation}
which implies that $W^{11}_{IJ}$ vanishes as well.
\vskip .2cm
\noindent $\bullet$ {\bf $\lambda=-1$.} For this case, which yields classically Lorentz invariant theories 
only for compact gaugings, $L^{01}_{IJ}$ vanishes while $L^{11}_{IJ}$ can be recast using (\ref{compgau}) in the form
\begin{equation}
L^{11}_{IJ} = -\frac{\lambda^2-1}{4} {\cal T}_{IA\hat{B}} {\cal T}_J^{\phantom{J}AB} =0 \ .
\end{equation}
For the Weyl anomaly, use of (\ref{HHetaeta}) shows that $W^{00}_{IJ}$ and $W^{01}_{IJ}$ trivially vanish while $W^{11}_{IJ}$ is rewritten as
\begin{equation}
W^{11}_{IJ} = - \frac{\lambda^2-1}{4} {\cal T}_{IAB} {\cal T}_J^{\phantom{J}AB} = 0\ .
\end{equation}
Hence the Lorentz and Weyl anomalies identically vanish.
\vskip .2cm
 
The above statements constitute our main results for the Lorentz and Weyl anomalies. For general gaugings with $\lambda=1$, conformal invariance dictates that the quantity $W^{11}_{IJ}$ given in (\ref{w11TDT}) must vanish; the meaning of this condition in the context of ${\cal N}=4$ gauged supergravity will be developed in \S\ref{corr}. For compact gaugings with $\lambda=\pm 1$ the Weyl anomaly is automatically vanishing. This 
leads us to suspect that all such models are actually WZW models, which are known to be
conformally invariant to all orders.  We will indeed establish the relation between
compact gaugings  and WZW models in
 \S\ref{wzw}.
 
Some remarks are in order. First, we emphasize that there was no need to resort to the classical equations of motion to verify that the Lorentz anomaly vanishes and that the Weyl anomaly takes the simple form (\ref{w11TDT}). Second, we note that equations with similar form to our final expressions (\ref{w11TDT0}) and (\ref{w11TDT}) for the Weyl anomaly have appeared earlier in the literature of WZW models \cite{Tseytlin:1993hm}. Finally, we observe that, had we imposed that the target space of our worldsheet theory was a local group manifold with the generalized three-form flux given by (\ref{htplussign}) or (\ref{htminussign}) from the very beginning, our background field expansion would simplify considerably and the various terms could be easily arranged using connections and curvatures with torsion as in \cite{Braaten:1985is,Mukhi:1985vy}. However, with an outlook towards
other potential applications of the interacting chiral boson model, we chose to present the expansion and the equations for the Weyl and Lorentz anomalies for the most general case and to specialize to the TDT case only in the preceding discussion. 

\subsection{Correspondence with \boldmath{${\cal N}=4$} gauged supergravity}
\label{corr}

If the sigma-models under consideration provide a worldsheet description of ${\cal N}=4$ gauged supergravity (for the case of electric gaugings), the condition for conformal invariance stated above should be equivalent to the condition that the gauged supergravity, with the corresponding embedding tensor, has a vacuum. Below, we shall prove that this equivalence does indeed hold.

For the analysis, it is useful to introduce some notation. We first consider the tensors
\begin{equation}
P_\pm^{ABCD}(H) = {1 \over 2} (\eta^{A(C} \eta^{D)B} \pm H^{A(C} H^{D)B}) \ .
\label{projectors}
\end{equation}
These are projection operators since, by virtue of the ${\rm O}(d,d)$ constraint (\ref{tanh}), they satisfy
\begin{equation}
P_\pm^{AB}{}_{CD}(H) P_\pm^{CDEF}(H) = P_\pm^{ABEF}(H)\ , \qquad P_\pm^{AB}{}_{CD}(H) P_\mp^{CDEF}(H) = 0\ ,
\end{equation}
and they also obey the completeness relation
\begin{equation}
P_+^{ABCD}(H)+P_-^{ABCD}(H) = \eta^{A(C} \eta^{D)B} \ .
\end{equation}
We also define the quantity
\begin{equation}
Z_{AB}(H) = {1 \over 2} ( \eta^{C[C'} \eta^{D']D} - H^{C[C'} H^{D']D} ) {\cal T}_{CDA} {\cal T}_{C'D'B}\ ,
\end{equation}
involving the antisymmetrized counterpart of the operator $P_-$ in (\ref{projectors}).

Using the above notation, the Weyl anomaly and the minimization condition for the supergravity potential can be written in a very useful form. Starting from the Weyl anomaly, we can rewrite (\ref{w11TDT}) compactly as
\begin{equation}
W_{AB} = P_{-AB}^{\phantom{-AB}CD}(H) Z_{CD}(H)\ ,
\label{compactweyl}
\end{equation}
where the superscript $11$ will be henceforth omitted. Turning to the gauged supergravity side, the derivative of the potential (\ref{gaugedsugpo}) is given by
\begin{equation}
{\partial V \over \partial M^{AB}} = - {1 \over 2} ( \eta^{CC'} \eta^{DD'} - M^{CC'} M^{DD'} ) {\cal T}_{ACD} {\cal T}_{BC'D'} = - Z_{AB}(M)\ ,
\label{dVdM}
\end{equation}
and hence involves the same structure appearing in eq.~(\ref{compactweyl}) for the Weyl anomaly, but for the presence of  the extra projector $P_-$ in the latter equation. 

The apparent discrepancy is easily resolved by noting that the minimization of the potential (\ref{gaugedsugpo}) is actually a constrained minimization problem due to the condition (\ref{lor}) imposed on $M$. Usually, such a minimization is performed by choosing a suitable ansatz for $M$ that respects (\ref{lor}) and varying the potential with respect to the parameters contained in the ansatz. However, since here we wish to consider a generic $M$ subject to (\ref{lor}), we are led to enforce this constraint by means of a Lagrange multiplier tensor $\Lambda^{AB}$ instead. Therefore, the quantity to be minimized is the modified potential
\begin{equation}
\hat{V} = V + \Lambda_A^{\phantom{A}B} ( M^{AC} \eta_{CB} - \eta^{AC} M_{CB} )\ .
\label{const}
\end{equation}
whose derivative equals
\begin{eqnarray}
{\partial \hat{V} \over \partial M^{AB}} &=& - Z_{AB}(M) + {1 \over 2} ( \eta_{AC} \eta_{BD} + M_{AB} M_{CD} ) \Lambda^{CD}\nonumber\\ 
&=& - Z_{AB}(M) + P_{+ABCD}(M) \Lambda^{CD}\ .
\label{dVhatdM}
\end{eqnarray}

Using eqs.~(\ref{compactweyl}) and (\ref{dVhatdM}), it is easy to prove the equivalence of the vanishing of the Weyl anomaly with the minimization of the supergravity potential at $M^{AB}=H^{AB}$. Notice that we are allowed to perform this identification since both are symmetric ${\rm O}(d,d)$ matrices. First, suppose that for the given gauging we have a vacuum, {\it i.e.}~there exists a Lagrange multiplier $\Lambda^{AB}$ such that ${\partial \hat{V} \over \partial M^{AB}} \bigr|_{M=H} = 0$ is satisfied. Then we have
\begin{equation}
Z_{AB}(H) - P_{+ABCD}(H) \Lambda^{CD} = 0\ .
\end{equation}
Acting on this equation with $P_-(H)$, we find
\begin{equation}
P_-^{ABCD}(H) Z_{CD}(H) = 0\ ,
\end{equation}
{\it i.e.}~the precise condition for the vanishing of the Weyl anomaly. 

Conversely, suppose that for a given gauging the Weyl anomaly vanishes,
\begin{equation}
P_{-}^{ABCD}(H) Z_{CD}(H) = 0\ .
\end{equation}
Then, completeness of the projection operators (\ref{projectors}) implies then that $Z_{AB}(H)$ lies in the invariant subspace of $P_+(H)$,
\begin{equation}
Z_{AB}(H) = P_{+ABCD}(H) Z^{CD}(H)\ .
\end{equation}
Plugging this into (\ref{dVhatdM}), we find
\begin{equation}
{\partial \hat{V} \over \partial M^{AB}}\biggr|_{M=H} = P_{+ABCD}(H) \left( \Lambda^{CD} - Z^{CD}(H) \right) \ .
\label{dVhatdM2}
\end{equation}
Then, setting $\Lambda^{AB}=Z^{AB}(H)$, we see that $M^{AB}=H^{AB}$ is a solution to the constrained minimization of the potential. Note that if the RHS of (\ref{dVhatdM2}) had the form $P_{+ABCD}(H) \Lambda^{CD} - Y_{AB}$ with generic $Y_{AB}$, it would not be possible
to invert it in general and 
yield a solution for $\Lambda^{AB}$. This completes the proof of equivalence.  
     
\subsection{Compact gaugings and WZW models}
\label{wzw}
 
The class of compact gaugings, {\it i.e.}~those described by structure constants that in addition to (\ref{structureconstantsodd}) satisfy (\ref{compgau}), are particularly interesting. As we have already discussed in subsection 4.3, the corresponding TDT  sigma-models
are (classically and quantum-mechanically) Lorentz invariant 
for two choices of generalized fluxes, namely $\lambda=\pm1$ in (\ref{lahtplussign}).
Moreover, they have vanishing Weyl divergence and therefore are conformally invariant.
This result begs for a deeper explanation and it is the main purpose of this subsection to provide it. 

If we impose the compact gauging condition (\ref{compgau}) on the 
decomposition  (\ref{thqr}) of the embedding tensor in terms of fluxes, we obtain
the following relations
\begin{equation}
H_{abc} = Q_{a}^{\phantom{a}bc}\ , \;\;\; R^{abc} = \tau_{bc}^{\phantom{bc}a} \ , \;\;\;
 \tau_{ab}^{\phantom{ab}c}  = -  \tau_{ac}^{\phantom{ac}b}\ , \;\;\;
 Q_c^{\phantom{c}ab} = - Q_a^{\phantom{a}cb}\ .
 \label{comgauexp}
\end{equation}
Notice that since $H_{abc}$ and $R^{abc}$ are fully antisymmetric, the
first two conditions enforce automatically the other two, {\it i.e.}~full antisymmetry of
$\tau_{ab}^{\phantom{ab}c} $ and $ Q_c^{\phantom{c}ab}$.  

For gaugings that satisfy these conditions, there is a particularly illuminating change of basis
of the gauge algebra generators  from $\{Z_a,X^a\}$  to $\{T_a^\pm = \frac{1}{2}(Z_a\pm X^a)\}$. It is straightforward to see that the gauge algebra
(\ref{thqr}) in terms of the new basis is a direct sum of algebras:
\begin{equation}
\text{[}T_a^\pm , T_b^\pm] = f_{ab}^{\pm c} T_c^\pm  \ ,\qquad
\text{[}T_a^+ , T_b^-] = 0\ ,
\end{equation}     
where the structure constants are 
$ f_{ab}^{\pm c} = \tau_{ab}^{\phantom{ab}c} \pm  Q_a^{\phantom{a}bc}$. Since the fluxes $\tau_{ab}^{\phantom{ab}c} $ and $ Q_c^{\phantom{c}ab}$ are fully antisymmetric, with indices raised and lowered with the $d$-dimensional Kronecker delta, the same is true for $f_{ab}^{\pm c}$. Therefore, the total algebra spanned by $T^\pm_a$ corresponds to the direct product of two compact and semisimple groups\footnote{An obvious exception to this is the special case 
where at least one set of the structure constants $ f_{ab}^{\pm c}$ vanishes, so that at least one of the factor groups is abelian.}
 ${\cal G}_L \times {\cal G}_R$. The latter are embedded in the maximal compact subgroup ${\rm O}(d) \times
{\rm O}(d)$ of the duality group ${\rm O}(d,d)$.  

It had been suggested in  \cite{Dabholkar:2005ve}, by comparing gauge symmetries and the moduli potential, that the compact gauging ${\rm SU}(2) \times {\rm SU}(2)$ (for $d=3$) should admit a string theory description in terms of a ${\rm SU}(2)$ WZW model. This proposal was proven to be correct in \cite{Dall'Agata:2008qz}, where it was explicitly shown that the TDT sigma-model for this gauging is actually the ${\rm SU}(2)$ WZW model in disguise.
We will now obtain a general result, that was already anticipated in \cite{Dall'Agata:2008qz}: for all compact gaugings ${\cal G}_L \times {\cal G}_R$ the TDT sigma-model is a sum of a chiral WZW model based on the group ${\cal G}_L$ and an antichiral one based on ${\cal G}_R$. This result will also provide the explanation for our starting observation, {\it i.e.}~that all compact gaugings lead to conformal models. 
      
We choose to parametrize the group element of the TDT for the case of a compact gauging
as
\begin{equation}
g = e^{\frac{1}{4} (y^a+\tilde y_a)(Z_a + X^a)}e^{\frac{1}{4} (y^a-\tilde y_a)(Z_a- X^a)}\ .
\end{equation}      
Defining the chiral and antichiral coordinates $y_L^i = y^i + \tilde y_i, y_R=y^i-
\tilde y_i$  and using the diagonal basis $T^\pm_a$ renders manifest the
product structure of the TDT:
\begin{equation}
g =  e^{y^i_L T^+_i} e^{y^i_R T^-_i} \equiv g_L g_R \ .
\end{equation}
The corresponding Maurer--Cartan form is decomposable $g^{-1} dg = 
g_L^{-1} dg_L + g^{-1}_R dg_R = L^a_i T^+_a dy_L^i + R^a_i T^-_a dy_R^i.$
Comparing this expression with the 
general form of the TDT vielbein  
$E^A_I\mathbb{X}_A   d\mathbb{Y}^I = g^{-1} dg$ we obtain
\begin{equation}
E^A_I = \frac{1}{2} \left(
\begin{array}{cc} 
L_i^a+R_i^a & L_i^a-R_i^a  \\
L_i^a-R_i^a & L_i^a+R_i^a 
\end{array} \right)\ .
\end{equation}
We can now construct $\eta_{IJ}$ and $H_{IJ}$  from (\ref{etaspec}) and  (\ref{hspec}),
where for simplicity we select $H_{AB} =\delta_{AB}$. We find that
\begin{equation}
H_{IJ} = \frac{1}{2} \left(
\begin{array}{cc} 
G^L_{ij} + G^R_{ij} & G^L_{ij}-G^R_{ij}  \\
G^L_{ij} -G^R_{ij} & G^L_{ij} + G^R_{ij}
\end{array} \right)\ , \quad
\eta_{IJ} = \frac{1}{2} \left(
\begin{array}{cc} 
G^L_{ij} - G^R_{ij} & G^L_{ij}+G^R_{ij}  \\
G^L_{ij}+G^R_{ij} & G^L_{ij} - G^R_{ij}
\end{array} \right)\ , 
\end{equation}
with $G^L_{ij} = L_i^a L_j^a$ and $ G^R_{ij} = R^a_i R^a_j$ being the bi-invariant metrics for the groups ${\cal G}_L$ and ${\cal G}_R$ respectively. Inserting these data, along with the generalized 3-form flux which reads
\begin{equation}
G_3 = d C_2 = \frac{\lambda}{2} ( f^{+c}_{ab} L^a L^b L^c - f^{-c}_{ab} R^a R^b R^c )\ ,
\end{equation}
in the interacting chiral boson Lagrangian (\ref{icb}), switching to the chiral field basis $y^i_{L,R}$ and introducing the usual chiral derivatives $\partial_\pm = \partial_0 \pm \partial_1$, brings the action to the form
\begin{equation}
S= \frac{1}{4} \int d^2 \sigma \left( (G^L_{ij} + \lambda B^{L}_{ij}) \partial_- y^i_L \partial_1 y^j_L
- (G^R_{ij} + \lambda B^{R}_{ij}) \partial_+ y^i_R \partial_1 y^j_R
\right)\ .
\end{equation}
The antisymmetric tensors $B_{ij}^{L,R}$ are the two-form potentials giving rise to the invariant NS-NS three-form fluxes for each group manifold, {\it i.e.}
\begin{equation}
dB^L = f^{+c}_{ab} L^a L^b L^c\ , \quad dB^R = f^{-c}_{ab} R^a R^b R^c \ .
\end{equation}

For any of the two choices $\lambda=\pm 1 $ that lead to a Lorentz invariant and conformal
chiral boson model,
the theory in this form is recognized to be the sum of a chiral WZW model for the group ${\cal G}_L$ and an antichiral one for ${\cal G}_R$ \cite{Sonnenschein:1988ug, Tseytlin:1990va}. Therefore, it is expected to be a conformal field theory to all orders in perturbation theory, in line with our finding that the corresponding TDT sigma-model is conformal for all compact gaugings. If the two groups are the same, for example when the $Q$-flux is zero and we have  $ f_{ab}^{+ c} =  f_{ab}^{- c} $, one can write the theory solely in terms of the fields $y^i$, therefore obtaining an ordinary, non-chiral, WZW model. The particular case ${\cal G}_L={\cal G}_R={\rm SU}(2)$ has been  originally analyzed in  \cite{Dall'Agata:2008qz} and it was shown to yield the ${\rm SU}(2)$ WZW model. 

For all compact gaugings,
the value of the potential (\ref{gaugedsugpo}) is negative 
\begin{equation}
 V = - \frac{4}{3} ( H_{abc} H^{abc} + R_{abc} R^{abc} )\ ,
\end{equation}
and therefore the spacetime of the effective supergravity theory is an ${\rm AdS}$ space. This is in accordance with the fact that without a non-trivial dilaton field we can embed a compact WZW model in a full-fledged string theory background only if some ${\rm AdS}$ factors are included. For example, a well-known background of this type is ${\rm AdS}_3 \times {\rm S}^3$ 
with NS-NS fluxes,
which  corresponds to a product ${\rm SL}(2,\mathbb{R}) \times  {\rm SU}(2)$ WZW model and describes the near-horizon region of a system of NS5-branes and fundamental strings.
  
The more general case where ${\cal G}_L$ is not the same as ${\cal G}_R$ is an asymmetric string compactification and since the worldsheet theory is under control, it provides a particularly tractable class of non-geometric string backgrounds that are worth to be studied further.

\subsection{Examples}

To give a concrete illustration of the use of eq.~(\ref{w11TDT}) for the Weyl anomaly and to make contact with known results, we will analyze some relatively simple gaugings. Considering the cases where only one or two types of fluxes are turned on, we will give explicit expressions for the quantity
\begin{equation}
Z_{AB} = {1 \over 2} ( \eta^{C[C'} \eta^{D']D} - H^{C[C'} H^{D']D} ) {\cal T}_{CDA} {\cal T}_{C'D'B}\ ,
\end{equation}
and for the tangent space components of the Weyl anomaly, defined by
\begin{equation}
W_{AB} = W_{ab} \delta^a_A \delta^b_B + W^{ab} \delta_{Aa} \delta_{Bb} + W_a^{\phantom{a}b} \delta^a_A \delta_{Bb} + W^a_{\phantom{a}b} \delta_{Aa} \delta^b_B \ .
\end{equation}
If the fluxes for a given gauging can be chosen so that $W_{AB}=0$, the associated models are conformal and correspond
to minima of the supergravity potential. Furthermore, since $Z_{AB}$ vanishes for a compact gauging, conformal models with nonzero $Z_{AB}$ correspond necessarily to non-compact gaugings with non-semisimple groups.

To avoid potential confusion, some remarks are in order. First, the expressions given below are derived using the tangent space basis where $H_{AB}$ and $\eta_{AB}$ take the form (\ref{freecbm}) and the decomposition assumes the form (\ref{thqr}), as is most natural for the TDT under consideration. The indices of the various types of fluxes appearing in that decomposition
are raised or lowered from their ``natural'' position using the $d$-dimensional Kronecker delta. If one wishes to work instead in a basis where $H_{AB}$ and $\eta_{AB}$ assume a different form (still satisfying (\ref{tanh}) of course), the expressions given below are not directly applicable and one is instructed to directly evaluate (\ref{compactweyl}) in the desired basis. Second, the various fluxes are not arbitrary, but are restricted by the Jacobi identity. For the examples with two types of flux, the non-trivial parts of the Jacobi identity will be explicitly stated.

\subsubsection{One type of flux}

As a warm-up, we first consider the simple cases where only one out of the four possible types of flux is turned on. For these configurations, we recover well-known results.
\vskip .2cm
\noindent $\bullet$ {\bf \boldmath{$H$}-flux.} We start with the  case of the NS-NS $H$-flux for which $Z_{AB}$ is given by 
\begin{equation}
Z_{AB} = - {1 \over 2} H_{cda} H^{cd}_{\phantom{cd}b} \delta^a_A \delta^b_B \ 
\end{equation}
while the Weyl anomaly reads
\begin{equation}
W_{ab} = - {1 \over 4} H_{cda} H^{cd}_{\phantom{cd}b}\ , \qquad W^{ab} = {1 \over 4} H_{cd}^{\phantom{cd}a} H^{cdb}\ .
\end{equation}
These are non-vanishing for any choice of $H_{abc}$, as expected.

\vskip .2cm
\noindent $\bullet$ {\bf \boldmath{$\tau$}-flux.} Passing to the case of geometric $\tau$-flux, we find that $Z_{AB}$ reads
\begin{equation}
Z_{AB} = - \tau_{ca}^{\phantom{ca}d} (\tau_{db}^{\phantom{db}c} + \tau^c_{\phantom{c}bd}) \delta^a_A \delta^b_B - {1 \over 2} \tau_{cd}^{\phantom{cd}a} \tau^{cdb} \delta_{Aa} \delta_{Bb} \ ,
\end{equation}
and, therefore, is generally nonzero. The Weyl anomaly reads
\begin{eqnarray}
W_{ab} &=& {1 \over 4} \left( \tau_{cda} \tau^{cd}_{\phantom{cd}b} - 2 \tau_{ca}^{\phantom{ca}d} (\tau_{db}^{\phantom{db}c}  + \tau^c_{\phantom{c}bd}) \right) \ , \nonumber\\
W^{ab} &=& - {1 \over 4} \left( \tau_{cd}^{\phantom{cd}a} \tau^{cdb} - 2 \tau_{c}^{\phantom{c}ad} (\tau_{d}^{\phantom{d}bc} + \tau^{cb}_{\phantom{cb}d}) \right) \ .
\end{eqnarray}
This is the familiar Scherk--Schwarz twisted torus compactification: $W_{ab}$ and $W^{ab}$ are just multiples of the derivative of the Scherk--Schwarz potential, evaluated at $G_{ab}=\delta_{ab}$. For the case of a semisimple group, the second term of each line drops out while the first term is proportional to the Cartan-Killing metric and is non-vanishing. However, for non-semisimple groups there may exist conformal models ({\it i.e.}~minima of the potential), the standard examples being provided by flat groups. 

As a concrete illustration, let us consider the TDT realization of the six-dimensional ($d=3$) flat group, whose structure is encoded in the following antisymmetric ${\cal T}$-structure constants
\begin{equation}
{\cal T}_{13\tilde{2}} = - N\ , \qquad {\cal T}_{12\tilde{3}} = N\ , \qquad \hbox{and cyclic}\ ,
\label{Tflatgroup}
\end{equation}
or, equivalently, in terms of the decomposition (\ref{thqr}),
\begin{equation}
\tau_{12}^{\phantom{12}3} = N\ , \quad \tau_{13}^{\phantom{13}2} = - N\ , \quad \tau_{21}^{\phantom{12}3} = - N\ , \quad \tau_{31}^{\phantom{13}2} = N\ .
\label{tauflatgroup}
\end{equation}
It is easy to check that in this case $Z_{AB}$ is indeed non-vanishing,
\begin{equation}
Z_{AB} = - N^2 ( \delta^2_A \delta^2_B + \delta^3_A \delta^3_B + \delta_{A2} \delta_{B2} + \delta_{A3} \delta_{B3} )\ .
\end{equation}
However, calculating the Weyl anomaly we find that it vanishes,
\begin{equation}
W_{ab} = W^{ab} = 0\ ,
\end{equation}
as expected from both the gravity and sigma-model viewpoints.

\vskip .2cm
\noindent $\bullet$ {\bf \boldmath{$Q$}-flux.} Next, we consider the case of the locally geometric $Q$-flux. Now, we have 
\begin{equation}
Z_{AB} = - {1 \over 2} Q_a^{\phantom{a}cd} Q_{bcd} \delta^a_A \delta^b_B  - Q_c^{\phantom{c}da} ( Q_d^{\phantom{d}cb} + Q^{c \phantom{d} b}_{\phantom{c}d} ) \delta_{Aa} \delta_{Bb} \ ,
\end{equation}
and
\begin{eqnarray}
W_{ab} &=& - {1 \over 4} \left( Q_a^{\phantom{a}cd} Q_{bcd} - 2 Q_{c\phantom{d} a}^{\phantom{c}d} ( Q_{d\phantom{c}b}^{\phantom{d}c} + Q^{c \phantom{d}}_{\phantom{c}db} ) \right) \ , \nonumber\\
W^{ab} &=& {1 \over 4} \left( Q^{acd} Q^b_{\phantom{b}cd} - 2 Q_c^{\phantom{c}da} ( Q_d^{\phantom{d}cb} + Q^{c \phantom{d} b}_{\phantom{c}d} ) \right) \ .
\end{eqnarray}
The analysis proceeds in an entirely analogous manner to the case of $\tau$-flux.

\vskip .2cm
\noindent $\bullet$ {\bf \boldmath{$R$}-flux.} Finally, we turn to the non-geometric $R$-flux. Now, we have
\begin{equation}
Z_{AB} = - {1 \over 2} R^{cda} R_{cd}^{\phantom{cd}b} \delta_{Aa} \delta_{Bb} 
\end{equation}
and
\begin{equation}
W_{ab} = {1 \over 4} R^{cd}_{\phantom{cd}a} R_{cdb}\ , \qquad W^{ab} = - {1 \over 4} R^{cda} R_{cd}^{\phantom{cd}b}\ .
\end{equation}
As for the case of $H$-flux, the Weyl anomaly is non-vanishing for any choice of $R^{abc}$.

\subsubsection{Two types of flux}

We study now the slightly more involved case where two types of flux are turned on. Examining all possible combinations, we recover the solutions corresponding to the compact gaugings of \S\ref{wzw} and we find a few other solutions to the Weyl invariance condition.
\vskip .2cm
\noindent $\bullet$ {\bf \boldmath{$H$}-flux and \boldmath{$\tau$}-flux.} The case where $H$- and $\tau$-fluxes are turned on is the one considered in ref.~\cite{Kaloper:1999yr}. We have
\begin{eqnarray}
Z_{AB} &=& - {1 \over 2} \left( H_{cda} H^{cd}_{\phantom{cd}b} + 2 \tau_{ca}^{\phantom{ca}d} (\tau_{db}^{\phantom{db}c} + \tau^c_{\phantom{c}bd}) \right) \delta^a_A \delta^b_B - {1 \over 2} \tau_{cd}^{\phantom{cd}a} \tau^{cdb} \delta_{Aa} \delta_{Bb}\nonumber\\ 
&-& {1 \over 2} H_{cda} \tau^{cdb} \delta^a_A \delta_{Bb} - {1 \over 2} \tau_{cd}^{\phantom{cd}a} H^{cd}_{\phantom{cd}b} \delta_{Aa} \delta^b_B\ ,
\end{eqnarray}
and
\begin{eqnarray}
W_{ab} &=& {1 \over 4} \left( \tau_{cda} \tau^{cd}_{\phantom{cd}b} - H_{cda} H^{cd}_{\phantom{cd}b} - 2 \tau_{ca}^{\phantom{ca}d} (\tau_{db}^{\phantom{db}c} + \tau^c_{\phantom{c}bd}) \right)\ , \nonumber\\
W^{ab} &=& - {1 \over 4} \left( \tau_{cd}^{\phantom{cd}a} \tau^{cdb} - H_{cd}^{\phantom{cd}a} H^{cdb} - 2 \tau_{c}^{\phantom{c}ad} (\tau_{d}^{\phantom{d}bc} + \tau^{cb}_{\phantom{cb}d}) \right)\ , \nonumber\\
W_a^{\phantom{a}b} &=& {1 \over 4} ( \tau_{cda} H^{cdb} - H_{cda} \tau^{cdb} )\ , \nonumber\\
W^a_{\phantom{a}b} &=& {1 \over 4} ( H_{cd}^{\phantom{cd}a} \tau^{cd}_{\phantom{cd}b} - \tau_{cd}^{\phantom{cd}a} H^{cd}_{\phantom{cd}b} )\ .
\end{eqnarray}
The Jacobi identities applying to this case are the standard $\tau\tau$ identity and
\begin{equation}
H_{e[ab} \tau_{c]d}^{\phantom{c]d}e} + \tau_{[ab}^{\phantom{[ab}e} H_{c]de} = 0\ .
\end{equation}
For fully antisymmetric $\tau$ with $\tau_{ab}^{\phantom{ab}c} = \pm H_{abc}$, the Weyl anomaly vanishes and the Jacobi identities are satisfied, while $Z_{AB}$ is nonzero. Such configurations of fluxes correspond necessarily to non-compact gaugings.

\vskip .2cm
\noindent $\bullet$ {\bf \boldmath{$H$}-flux and \boldmath{$Q$}-flux.} When only $H$- and $Q$-fluxes are turned on, we have
\begin{equation}
Z_{AB} =  - Q_c^{\phantom{c}da} ( Q_d^{\phantom{d}cb} + Q^{c \phantom{d} b}_{\phantom{c}d} ) \delta_{Aa} \delta_{Bb} - {1 \over 2} ( H_{cda} - Q_{acd} ) ( H^{cd}_{\phantom{cd} b} - Q_b^{\phantom{b}cd} ) \delta^a_A \delta^b_B 
\end{equation}
and
\begin{eqnarray}
W_{ab} &=& - {1 \over 4} \left( ( H_{cda} - Q_{acd} ) ( H^{cd}_{\phantom{cd} b} - Q_b^{\phantom{b}cd} ) - 2 Q_{c\phantom{d}a}^{\phantom{c}d} ( Q_{d\phantom{c}b}^{\phantom{d}c} + Q^{c}_{\phantom{c}db} ) \right)\ , \nonumber\\
W^{ab} &=& {1 \over 4} \left( ( H_{cd}^{\phantom{cd}a} - Q^a_{\phantom{a}cd} ) ( H^{cdb} - Q^{bcd} ) - 2 Q_c^{\phantom{c}da} ( Q_d^{\phantom{d}cb} + Q^{c \phantom{d} b}_{\phantom{c}d} ) \right)\ .
\end{eqnarray}
The relevant Jacobi identities are the $QQ$ identity and
\begin{equation}
H_{e[ab} Q_{c]}^{\phantom{c]}de} = 0\ .
\end{equation}
For fully antisymmetric $Q$ with $Q_c^{\phantom{c}ab} = H_{abc}$, the Weyl anomaly and $Z_{AB}$ vanish and the Jacobi identities are satisfied. This is actually expected since, as we saw in \S\ref{wzw}, fluxes satisfying this condition correspond necessarily to compact gaugings which lead always to conformal models.

\vskip .2cm
\noindent $\bullet$ {\bf \boldmath{$H$}-flux and \boldmath{$R$}-flux.} In the presence of $H$- and $R$-fluxes, we have
\begin{equation}
Z_{AB} = - {1 \over 2} H_{cda} H^{cd}_{\phantom{cd} b} \delta^a_A \delta^b_B - {1 \over 2} R^{cda} R_{cd}^{\phantom{cd} b} \delta_{Aa} \delta_{Bb}
+ {1 \over 2} H_{cda} R^{cdb} \delta^a_A \delta_{Bb} + {1 \over 2} R^{cda} H_{cdb} \delta_{Aa} \delta^b_B 
\end{equation}
and
\begin{eqnarray}
W_{ab} &=& - {1 \over 4} (H_{cda} H^{cd}_{\phantom{cd} b} - R_{cda} R^{cd}_{\phantom{cd} b} )\ , \nonumber\\
W^{ab} &=& {1 \over 4} ( H^{cda} H_{cd}^{\phantom{cd} b} - R^{cda} R_{cd}^{\phantom{cd} b} )\ , \nonumber\\
W_a^{\phantom{a}b} &=& {1 \over 4} (H_{cda} R^{cdb} - R^{cd}_{\phantom{cd}a} H_{cd}^{\phantom{cd}b} ) \ , \nonumber\\
W^a_{\phantom{a}b} &=& - {1 \over 4} ( H_{cd}^{\phantom{cd}a} R^{cdb} - R^{cda} H_{cdb} ) \ .
\end{eqnarray}
The Jacobi identity for this case is quite restrictive,
\begin{equation}
H_{abe} R^{cde} = 0\ .
\label{jachr}
\end{equation}
An obvious way to make the Weyl anomaly vanish is to take $H_{abc}=\pm R^{abc}$. However, in this case the Jacobi identity (\ref{jachr}) demands that the connection $H_{abc}$ be flat. 

\vskip .2cm
\noindent $\bullet$ {\bf Drinfel'd doubles: \boldmath{$\tau$}-flux and \boldmath{$Q$}-flux.} For configurations with $\tau$- and $Q$-fluxes, the gauge algebra falls into the category of Drinfel'd doubles, usually appearing in the context of Poisson--Lie T-duality. We find
\begin{eqnarray}
Z_{AB} &=& \left( \tau_{ca}^{\phantom{ca}d} (Q_d^{\phantom{d}cb} + Q^{c \phantom{d} b}_{\phantom{c}d} ) + {1 \over 2} Q_a^{\phantom{a}cd} \tau_{cd}^{\phantom{cd}b} \right) \delta^a_A \delta_{Bb} \nonumber\\
&+& \left( Q_c^{\phantom{c}da} (\tau_{db}^{\phantom{db}c} + \tau^c_{\phantom{c}bd}) + {1 \over 2} \tau_{cd}^{\phantom{cd}a} Q_b^{\phantom{b}cd} \right) \delta_{Aa} \delta^b_B \nonumber\\
&-& \left( \tau_{ca}^{\phantom{ca}d} (\tau_{db}^{\phantom{db}c} + \tau^c_{\phantom{c}bd}) + {1 \over 2} Q_a^{\phantom{a}cd} Q_{bcd} \right) \delta^a_A \delta^b_B \nonumber\\
&-& \left( Q_c^{\phantom{c}da} ( Q_d^{\phantom{d}cb} + Q^{c \phantom{d} b}_{\phantom{c}d}) + {1 \over 2} \tau_{cd}^{\phantom{cd}a} \tau^{cdb} \right) \delta_{Aa} \delta_{Bb} 
\end{eqnarray}
and
\begin{eqnarray}
W_{ab} &=& {1 \over 4} \left( \tau_{cda} \tau^{cd}_{\phantom{cd}b} - Q_a^{\phantom{a}cd} Q_{bcd} - 2 \tau_{ca}^{\phantom{ca}d} (\tau_{db}^{\phantom{db}c} + \tau^c_{\phantom{c}bd}) + 2 Q_{c\phantom{d}a}^{\phantom{c}d} ( Q_{d\phantom{c}b}^{\phantom{d}c} + Q^{c}_{\phantom{c}db}) \right)\ , \nonumber\\
W^{ab} &=& {1 \over 4} \left( Q^{acd} Q^b_{\phantom{b}cd} - \tau_{cd}^{\phantom{cd}a} \tau^{cdb} - 2 Q_c^{\phantom{c}da} ( Q_d^{\phantom{d}cb} + Q^{c \phantom{d} b}_{\phantom{c}d}) + 2 \tau_{c}^{\phantom{c}ad} (\tau_{d}^{\phantom{d}bc} + \tau^{cb}_{\phantom{cb}d}) \right)\ , \nonumber\\
W_a^{\phantom{a} b} &=& {1 \over 4} \left( Q_a^{\phantom{a}cd} \tau_{cd}^{\phantom{cd}b} - \tau_{cda} Q^{bcd}
 - 2 Q_{c\phantom{d}a}^{\phantom{c}d} (\tau_{d}^{\phantom{d}bc} + \tau^{cb}_{\phantom{cb}d}) + 2 \tau_{ca}^{\phantom{ca}d} (Q_d^{\phantom{d}cb} + Q^{c \phantom{d} b}_{\phantom{c}d} ) \right)\ , \nonumber\\
W^a_{\phantom{a}b} &=& {1 \over 4} \left( \tau_{cd}^{\phantom{cd}a} Q_b^{\phantom{b}cd} - Q^{acd} \tau_{cdb} - 2 \tau_{c}^{\phantom{c}ad} (Q_{d\phantom{c}b}^{\phantom{d}c} + Q^{c \phantom{d}}_{\phantom{c}db} ) + 2 Q_c^{\phantom{c}da} (\tau_{db}^{\phantom{db}c} + \tau^c_{\phantom{c}bd}) \right)\ .
\label{tauQWeyl}
\end{eqnarray}
The Jacobi identities are the standard $\tau\tau$ and $QQ$ identities plus
\begin{equation}
\tau_{ab}^{\phantom{ab}e} Q_e^{\phantom{e}cd} - 2 \tau_{e[a}^{\phantom{e[a}c} Q_{b]}^{\phantom{b]}de} + 2 \tau_{e[a}^{\phantom{e[a}d} Q_{b]}^{\phantom{b]}ce}= 0\ .
\label{jactq}
\end{equation}
The Weyl anomaly can vanish if we take fully antisymmetric $\tau$ and $Q$ with $\tau_{ab}^{\phantom{ab}c} = \pm Q_c^{\phantom{c}ab}$. However, in analogy to the previous case, the Jacobi identity (\ref{jactq}) combined with the first Bianchi identity for the Riemann tensor demands that the connection $\tau_{ab}^{\phantom{ab}c}$ be flat.

\vskip .2cm
\noindent $\bullet$ {\bf \boldmath{$\tau$}-flux and \boldmath{$R$}-flux.} In this case, we have
\begin{equation}
Z_{AB} = - \tau_{ca}^{\phantom{ca}d} (\tau_{db}^{\phantom{db}c} + \tau^c_{\phantom{c}bd}) \delta^a_A \delta^b_B - {1 \over 2} ( \tau_{cd}^{\phantom{cd}a} -  R_{cd}^{\phantom{cd}a} ) ( \tau^{cdb} - R^{cdb} ) \delta_{Aa} \delta_{Bb}
\end{equation}
and
\begin{eqnarray}
W_{ab} &=& {1 \over 4} \left( ( \tau_{cda} -  R_{cda} ) ( \tau^{cd}_{\phantom{cd}b} - R^{cd}_{\phantom{cd}b} ) - 2 \tau_{ca}^{\phantom{ca}d} (\tau_{db}^{\phantom{db}c} + \tau^c_{\phantom{c}bd}) \right)\ , \nonumber\\
W^{ab} &=& - {1 \over 4} \left( ( \tau_{cd}^{\phantom{cd}a} -  R_{cd}^{\phantom{cd}a} ) ( \tau^{cdb} - R^{cdb} ) - 2 \tau_{c}^{\phantom{c}ad} (\tau_{d}^{\phantom{d}bc} + \tau^{cb}_{\phantom{cb}d}) \right)\ .
\end{eqnarray}
The Jacobi identities are the standard $\tau\tau$ identity plus
\begin{equation}
\tau_{ae}^{\phantom{ae}[b} R^{cd]e} = 0\ .
\end{equation}
This setup is T-dual to the one with $H$- and $R$-fluxes. For fully antisymmetric $\tau$ with $\tau_{ab}^{\phantom{ab}c} = R^{abc}$, the Weyl anomaly and $Z_{AB}$ vanish and the Jacobi identities are satisfied. As in the case of equal $H$- and $Q$-flux, these configurations describe compact gaugings which are guaranteed to be conformal.

\vskip .2cm
\noindent $\bullet$ {\bf \boldmath{$Q$}-flux and \boldmath{$R$}-flux.} 
We find
\begin{eqnarray}
Z_{AB}&=& - {1 \over 2} Q_a^{\phantom{a}cd} Q_{bcd} \delta^a_A \delta^b_B - {1 \over 2} \left( R^{cda}R_{cd}^{\phantom{cd}b} + 2 Q_c^{\phantom{c}da} ( Q_d^{\phantom{d}cb} + Q^{c \phantom{d} b}_{\phantom{c}d} ) \right)\delta_{Aa} \delta_{Bb} \nonumber\\
&-& {1 \over 2} Q_a^{\phantom{a}cd} R^b_{\phantom{b}cd} \delta^a_A \delta_{Bb} - {1 \over 2} R^{acd} Q_{bcd} \delta_{Aa} \delta^b_B
\end{eqnarray}
and
\begin{eqnarray}
W_{ab} &=& - {1 \over 4} \left( Q_a^{\phantom{a}cd} Q_{bcd} - R^{cd}_{\phantom{cd}a} R_{cdb} - 2 Q_{c\phantom{d}a}^{\phantom{c}d} ( Q_{d\phantom{c}b}^{\phantom{d}c} + Q^{c}_{\phantom{c}db}) \right)\ , \nonumber\\
W^{ab} &=& {1 \over 4} \left( Q^{acd} Q^b_{\phantom{b}cd} - R^{cda}R_{cd}^{\phantom{cd}b} - 2 Q_c^{\phantom{c}da} ( Q_d^{\phantom{d}cb} + Q^{c \phantom{d} b}_{\phantom{c}d})  \right)\ , \nonumber\\
W_a^{\phantom{a} b} &=& {1 \over 4} \left( R_a^{\phantom{a}cd} Q^b_{\phantom{b}cd} - Q_a^{\phantom{a}cd} R^b_{\phantom{b}cd} \right)\ , \nonumber\\
W^a_{\phantom{a}b} &=& {1 \over 4} \left( Q^{acd} R_{bcd} - R^{acd} Q_{bcd} \right)\ .
\end{eqnarray}
In addition to the usual $QQ$ condition, the Jacobi identities also require
\begin{equation}
Q_e^{\phantom{e}[ab} R^{c]de} + R^{e[ab} Q_e^{\phantom{e}c]d} = 0\ .
\end{equation}
This case is T-dual to that  of $\tau$- and $H$-fluxes. For fully antisymmetric $Q$ with $Q_c^{\phantom{c}ab} = \pm R^{abc}$, the Weyl anomaly vanishes and the Jacobi identities are satisfied, while $Z_{AB}$ is nonzero. Such configurations correspond to non-compact gaugings with non-semisimple groups.

\section{Future directions}
\label{conclusions}

The results of this paper firmly establish that the interacting chiral boson models of ref.~\cite{Dall'Agata:2008qz} based on twisted doubled tori are the underlying worldsheet theories of electric gaugings of ${\cal N}=4$ supergravity. They also show that twisted doubled tori that satisfy the conformal invariance conditions are genuine string backgrounds and not just a convenient bookkeeping tool. The overall message is that doubled geometries provide a powerful  theoretical framework for a systematic study of non-geometric string compactifications.

The analysis presented in this paper can be extended into several interesting directions. One important extension is to include the dilaton by supplementing the interacting chiral boson model with the Fradkin--Tseytlin term. First, this is expected to complete the analogy between the Weyl anomaly of our TDT models and the equations of motion of the full scalar sector of ${\cal N}=4$ supergravity with electric gaugings. Second, the dilaton is also relevant if one wants to uncover more general backgrounds than the TDT that lead to ${\cal N}=4$ theories with gauging of the axionic shift symmetry \cite{Derendinger:2007xp}. Another generalization is to include non-chiral bosons $X^m$ corresponding to the non-compact spacetime coordinates; this can be carried out along the lines of \cite{Berman:2007xn} and is expected to lead to similar results. 

At a more conceptual level, one could further examine the conformal field theory aspects of interacting chiral boson models on TDT, and in particular their relation with WZW models
and deformations thereof. Apart from the equivalence of the TDT models based on compact gaugings to chiral WZW models, the formula for the Weyl anomaly in the general case also suggests a possible relation with certain deformations of the chiral gauged WZW models considered in \cite{Tseytlin:1993hm}. It would be interesting to put this relation on a firm footing and to provide an explicit construction of the corresponding 
conformal field theories.

On the other hand, our formalism is general and allows to investigate other situations besides the TDT discussed here. In our opinion, the most interesting such application is to consider backgrounds where $H_{IJ}$ has arbitrary dependence on the doubled coordinates while $\eta_{IJ}$ takes the form (\ref{freecbm}) and both are related as in (\ref{lorentz1}). In this case $H_{IJ}$ is in the coset ${{\rm O}(d,d) \over {\rm O}(d)\times {\rm O}(d)}$ and can be written as in (\ref{stansin}), with the metric $G_{ij}$ and the B-field $B_{ij}$ being ${\mathbb Y}$-dependent. In such a case, it is natural to conjecture that the interacting chiral boson model provides a worldsheet description of the theories discussed recently in \cite{Hull:2009mi}. In this paper, closed string field theory methods were employed
to construct an effective action up to cubic order in a linearized expansion on the doubled torus. If our conjecture is correct, the beta functional of the chiral boson models should provide us with the full nonlinear completion of the linear and quadratic equations of motion presented in \cite{Hull:2009mi}. Work in this direction is currently under way \cite{ICBandHZ}.

Finally, an important generalization would be to formulate the chiral boson models for other types of doubled geometries that are not twisted doubled tori. This question is tied to the challenging task of uncovering new Lorentz invariant theories based on such models. An even more exotic goal would be to construct models that realize the full duality group, {\it i.e.}~incorporating S-duality, and therefore underlie all types of gaugings, not only the electric ones. Such models, if they exist at all, would give valuable insights into the physics of general non-geometric string backgrounds.

\appendix

\section{Notations and conventions}

Regarding indices, our conventions are the following. On the doubled target space the coordinates are $\mathbb{Y}^I$, curved indices are labeled by $I,J,\ldots=1,\ldots,2d$ while tangent space indices are labelled by $A,B,\ldots=1,\ldots,2d$. These $2d$-dimensional indices will be split into $d$-dimensional indices in the basis where the ${\rm O}(d,d)$ invariant tensor $\eta_{AB}$ takes the form given in (\ref{etaspec}), and the decomposition reads $V^I = \{ V^i,V^{\tilde{i}} \} \equiv \{ V^i, \tilde{V}_i\}$ or $V^A = \{ V^a,V^{\tilde{a}} \} \equiv \{ V^a, \tilde{V}_a\}$. On the worldsheet, the coordinates are $\sigma^{\mu}=(\tau,\sigma)$, curved indices are labelled by $\mu,\nu,\ldots=0,1$ and tangent space indices are labelled by $\alpha,\beta,\ldots=0,1$. The flat worldsheet metric is $g_{\mu \nu}= g^{\mu \nu}= {\rm diag}(1,-1)$ and the $\epsilon$ tensor is defined according to $\epsilon^{\mu\nu}=-\epsilon^{\nu\mu}=-\epsilon_{\mu\nu}$ with $\epsilon^{01}=1$. For the symmetrization and antisymmetrization of indices we use the convention $T_{(AB)} = {1 \over 2} (T_{AB}+T_{BA})$ and $T_{[AB]} = {1 \over 2} (T_{AB}-T_{BA})$.

Regarding differential geometry we follow the standard conventions. Namely
\begin{eqnarray}
D_I V^A_J &=& \partial_I V^A_J +  \Omega_{I \phantom{A} B}^{\phantom{I} A} V^B_J -\Gamma_{IJ}^K V^A_K \ ,\nonumber\\
D_I V_A^J &=& \partial_I V_A^J -  \Omega_{I \phantom{A} A}^{\phantom{I} B} V_B^J +\Gamma^J_{IK} V^K_A\ ,
\nonumber
\end{eqnarray}
where $\Gamma^K_{IJ}$ are the usual Christoffel symbols while the torsionless spin connection is defined by
\begin{equation}
d E^A + \Omega^A_{\phantom{A}B} \wedge E^B = 0\ ,
\end{equation}
with $ \Omega^A_{\phantom{A}B} = \Omega_{I \phantom{A} B}^{\phantom{I} A} \, d{\mathbb Y}^I$.
The Ricci 2-form is
\begin{equation}
R^A_{\phantom{A} B} = d  \Omega^A_{\phantom{A}B} +  \Omega^A_{\phantom{A}C} \wedge  \Omega^C_{\phantom{C}B}\ ,
 \end{equation}
with the Riemann tensor being
$R^A_{\phantom{A}B} = \frac{1}{2} R^A_{\phantom{A} B IJ} d{\mathbb Y}^I  \wedge d{\mathbb Y}^J.$ Furthermore, 
the covariant constancy of the vielbein implies
\begin{equation}
D_I E^A_J = \partial_I E^A_J + \Omega_{I\phantom{A}C}^{\phantom{I}A} E^C_J - \Gamma_{IJ}^K E^A_K =0\ .
\end{equation}

\section*{Acknowledgments}

\noindent  We are indebted to G.~Dall'Agata for very helpful discussions as well as for his comments on the manuscript. We would like to thank M.~Blau, A.~Fotopoulos, K.~Sfetsos and K.~Siampos for helpful discussions and  D.~Thompson for useful correspondence. This work was partially supported by the Swiss National Science Foundation.

\bibliographystyle{plain}

\begin{thebibliography}{99}


\bibitem{Wecht:2007wu}
  B.~Wecht,
  ``Lectures on Nongeometric Flux Compactifications,''
  Class.\ Quant.\ Grav.\  {\bf 24}, S773 (2007)
  [arXiv:0708.3984 [hep-th]].


\bibitem{Cremmer:1977tt}
  E.~Cremmer, J.~Scherk and S.~Ferrara,
  ``SU(4) Invariant Supergravity Theory,''
  Phys.\ Lett.\  B {\bf 74}, 61 (1978).

\bibitem{Cremmer:1979up}
  E.~Cremmer and B.~Julia,
  ``The {\rm SO}(8) Supergravity,''
  Nucl.\ Phys.\  B {\bf 159}, 141 (1979).


\bibitem{deWit:2005ub}
  B.~de Wit, H.~Samtleben and M.~Trigiante,
  ``Magnetic charges in local field theory,''
  JHEP {\bf 0509} (2005) 016
  [arXiv:hep-th/0507289].

\bibitem{Shelton:2005cf} J.~Shelton, W.~Taylor and B.~Wecht, ``Nongeometric flux compactifications,'' JHEP {\bf 0510}, 085 (2005) [arXiv:hep-th/0508133].

\bibitem{Aldazabal:2006up}
  G.~Aldazabal, P.~G.~Camara, A.~Font and L.~E.~Ibanez,
  ``More dual fluxes and moduli fixing,''
  JHEP {\bf 0605}, 070 (2006)
  [arXiv:hep-th/0602089].

	
\bibitem{Shelton:2006fd} J.~Shelton, W.~Taylor and B.~Wecht, ``Generalized flux vacua,'' JHEP {\bf 0702}, 095 (2007) [arXiv:hep-th/0607015].

\bibitem{Hellerman:2002ax}
  S.~Hellerman, J.~McGreevy and B.~Williams,
  ``Geometric Constructions of Nongeometric String Theories,''
  JHEP {\bf 0401}, 024 (2004)
  [arXiv:hep-th/0208174].

\bibitem{Dabholkar:2002sy}
  A.~Dabholkar and C.~Hull,
  ``Duality twists, orbifolds, and fluxes,''
  JHEP {\bf 0309}, 054 (2003)
  [arXiv:hep-th/0210209].


\bibitem{Kachru:2002sk}
  S.~Kachru, M.~B.~Schulz, P.~K.~Tripathy and S.~P.~Trivedi,
  ``New supersymmetric string compactifications,''
  JHEP {\bf 0303}, 061 (2003)
  [arXiv:hep-th/0211182].
  
  
  
\bibitem{Witten:1988zd}
  E.~Witten,
  ``Space-time and Topological Orbifolds,''
  Phys.\ Rev.\ Lett.\  {\bf 61}, 670 (1988).
  
\bibitem{Duff:1989tf}
  M.~J.~Duff,
  ``Duality Rotations In String Theory,''
  Nucl.\ Phys.\  B {\bf 335}, 610 (1990).

\bibitem{Tseytlin:1990nb} A.~A.~Tseytlin, ``Duality Symmetric Formulation Of String World Sheet Dynamics,'' Phys.\ Lett.\ B {\bf 242} (1990) 163. 
	
\bibitem{Tseytlin:1990va} A.~A.~Tseytlin,``Duality Symmetric Closed String Theory And Interacting Chiral Scalars,'' Nucl.\ Phys.\ B {\bf 350} (1991) 395.

\bibitem{Maharana:1992my}
  J.~Maharana and J.~H.~Schwarz,
  ``Noncompact symmetries in string theory,''
  Nucl.\ Phys.\  B {\bf 390}, 3 (1993)
  [arXiv:hep-th/9207016].


\bibitem{Hull:2004in} C.~M.~Hull, ``A geometry for non-geometric string backgrounds,'' JHEP {\bf 0510} (2005) 065 [arXiv:hep-th/0406102].


\bibitem{Hull:2006va} C.~M.~Hull, ``Doubled geometry and T-folds,'' JHEP {\bf 0707} (2007) 080 [arXiv:hep-th/0605149].

\bibitem{Hull:2007zu} C.~M.~Hull, ``Generalised geometry for M-theory,'' JHEP {\bf 0707} (2007) 079 [arXiv:hep-th/0701203].

  
  
\bibitem{Dall'Agata:2007sr} G.~Dall'Agata, N.~Prezas, H.~Samtleben and M.~Trigiante, ``Gauged Supergravities from Twisted Doubled Tori and Non-Geometric String Backgrounds,'' Nucl.\ Phys.\ B {\bf 799}, 80 (2008) [arXiv:0712.1026 [hep-th]].
	
	
	
\bibitem{Hull:2007jy}
  C.~M.~Hull and R.~A.~Reid-Edwards,
  ``Gauge Symmetry, T-Duality and Doubled Geometry,''
  JHEP {\bf 0808}, 043 (2008)
  [arXiv:0711.4818 [hep-th]].


\bibitem{Scherk:1979zr} J.~Scherk and J.~H.~Schwarz, ``How To Get Masses From Extra Dimensions,'' Nucl.\ Phys.\ B {\bf 153} (1979) 61.


\bibitem{Kaloper:1999yr}
  N.~Kaloper and R.~C.~Myers,
  ``The O(dd) story of massive supergravity,''
  JHEP {\bf 9905}, 010 (1999)
  [arXiv:hep-th/9901045].
  
  
  
\bibitem{Siegel:1983es}
  W.~Siegel,
  ``Manifest Lorentz Invariance Sometimes Requires Nonlinearity,''
  Nucl.\ Phys.\  B {\bf 238}, 307 (1984).
  
  
  
\bibitem{Floreanini:1987as}
  R.~Floreanini and R.~Jackiw,
  ``Selfdual Fields As Charge Density Solitons,''
  Phys.\ Rev.\ Lett.\  {\bf 59}, 1873 (1987).



	
\bibitem{Dall'Agata:2008qz}
  G.~Dall'Agata and N.~Prezas,
  ``Worldsheet theories for non-geometric string backgrounds,''
  JHEP {\bf 0808} (2008) 088
  [arXiv:0806.2003 [hep-th]].



  
\bibitem{Albertsson:2008gq}
  C.~Albertsson, T.~Kimura and R.~A.~Reid-Edwards,
  ``D-branes and doubled geometry,''
  JHEP {\bf 0904}, 113 (2009)
  [arXiv:0806.1783 [hep-th]].



\bibitem{Hull:2009sg}
  C.~M.~Hull and R.~A.~Reid-Edwards,
  ``Non-geometric backgrounds, doubled geometry and generalised T-duality,''
  arXiv:0902.4032 [hep-th].
  
    
  
\bibitem{ReidEdwards:2009nu}
  R.~A.~Reid-Edwards,
  ``Flux compactifications, twisted tori and doubled geometry,''
  JHEP {\bf 0906}, 085 (2009)
  [arXiv:0904.0380 [hep-th]].


\bibitem{Halmagyi:2009te}
  N.~Halmagyi,
  ``Non-geometric Backgrounds and the First Order String Sigma Model,''
  arXiv:0906.2891 [hep-th].

		
\bibitem{Berman:2007xn} D.~S.~Berman, N.~B.~Copland and D.~C.~Thompson, ``Background Field Equations for the Duality Symmetric String,'' Nucl.\ Phys.\ B {\bf 791} (2008) 175 [arXiv:0708.2267 [hep-th]]. 


\bibitem{Berman:2007yf} D.~S.~Berman and D.~C.~Thompson, ``Duality Symmetric Strings, Dilatons and O(d,d) Effective Actions,'' Phys.\ Lett.\ B {\bf 662} (2008) 279 [arXiv:0712.1121 [hep-th]].  


\bibitem{Depireux:1988yi}
  D.~A.~Depireux, S.~J.~J.~Gates and Q.~H.~J.~Park,
  ``Lefton - Righton Formulation Of Massless Thirring Models,''
  Phys.\ Lett.\  B {\bf 224}, 364 (1989).

\bibitem{Gates:1987sy}
  S.~J.~J.~Gates and W.~Siegel,
  ``Leftons, Rightons, Nonlinear Sigma Models, and Superstrings,''
  Phys.\ Lett.\  B {\bf 206}, 631 (1988).


 
\bibitem{AlvarezGaume:1981hn}
  L.~Alvarez-Gaume, D.~Z.~Freedman and S.~Mukhi, ``The Background Field Method And The Ultraviolet Structure Of The Supersymmetric Nonlinear Sigma Model,''
  Annals Phys.\  {\bf 134}, 85 (1981).

\bibitem{Braaten:1985is}
  E.~Braaten, T.~L.~Curtright and C.~K.~Zachos,
  ``Torsion And Geometrostasis In Nonlinear Sigma Models,''
  Nucl.\ Phys.\  B {\bf 260}, 630 (1985).

\bibitem{Mukhi:1985vy}
  S.~Mukhi,
  ``The Geometric Background Field Method, Renormalization And The Wess-Zumino
  Term In Nonlinear Sigma Models,''
  Nucl.\ Phys.\  B {\bf 264}, 640 (1986).

\bibitem{Howe:1986vm}
  P.~S.~Howe, G.~Papadopoulos and K.~S.~Stelle,
  ``The Background Field Method And The Nonlinear Sigma Model,''
  Nucl.\ Phys.\  B {\bf 296}, 26 (1988).

\bibitem{Curtright:1984dz}
  T.~L.~Curtright and C.~K.~Zachos,
  ``Geometry, topology and supersymmetry in nonlinear models,''
  Phys.\ Rev.\ Lett.\  {\bf 53}, 1799 (1984).


\bibitem{Hull:2009mi}
  C.~Hull and B.~Zwiebach,
  ``Double field theory,''
  arXiv:0904.4664 [hep-th].
  
  
  
\bibitem{Gates:1982ct}
  S.~J.~J.~Gates and B.~Zwiebach,
  ``Gauged N=4 supergravity theory with a new scalar potential,''
  Phys.\ Lett.\  B {\bf 123}, 200 (1983).

\bibitem{de Roo:1984gd}
  M.~de Roo,
  ``Matter coupling In N=4 supergravity,''
  Nucl.\ Phys.\  B {\bf 255}, 515 (1985).

\bibitem{deRoo:1985jh}
  M.~de Roo and P.~Wagemans,
  ``Gauge matter coupling In N=4 supergravity,''
  Nucl.\ Phys.\  B {\bf 262}, 644 (1985).

\bibitem{deRoo:1986yw}
  M.~de Roo and P.~Wagemans,
  ``Partial supersymmetry breaking in N=4 supergravity,''
  Phys.\ Lett.\  B {\bf 177}, 352 (1986).

\bibitem{Bergshoeff:1985ms}
  E.~Bergshoeff, I.~G.~Koh and E.~Sezgin,
  ``Coupling of Yang-Mills To N=4, D=4 supergravity,''
  Phys.\ Lett.\  B {\bf 155}, 71 (1985).

\bibitem{Schon:2006kz}
  J.~Sch\"on and M.~Weidner,
  ``Gauged N = 4 supergravities,''
  JHEP {\bf 0605}, 034 (2006)
  [arXiv:hep-th/0602024].

\bibitem{Derendinger:2007xp}
  J.~P.~Derendinger, P.~M.~Petropoulos and N.~Prezas,
  ``Axionic symmetry gaugings in N = 4 supergravities and their
  higher-dimensional origin,''
  Nucl.\ Phys.\  B {\bf 785}, 115 (2007)
  [arXiv:0705.0008 [hep-th]].

\bibitem{ReidEdwards:2008rd}
  R.~A.~Reid-Edwards and B.~Spanjaard,
  ``N=4 gauged supergravity from duality-twist compactifications of string
  theory,''
  JHEP {\bf 0812}, 052 (2008)
  [arXiv:0810.4699 [hep-th]].

\bibitem{Aldazabal:2008zza}
  G.~Aldazabal, P.~G.~Camara and J.~A.~Rosabal,
  ``Flux algebra, Bianchi identities and Freed-Witten anomalies in F-theory
  compactifications,''
  Nucl.\ Phys.\  B {\bf 814}, 21 (2009)
  [arXiv:0811.2900 [hep-th]].

\bibitem{Dall'Agata:2009gv}
  G.~Dall'Agata, G.~Villadoro and F.~Zwirner,
  ``Type-IIA flux compactifications and N=4 gauged supergravities,''
  arXiv:0906.0370 [hep-th].


\bibitem{Tseytlin:1993hm}
  A.~A.~Tseytlin,
  ``On A 'Universal' class of WZW type conformal models,''
  Nucl.\ Phys.\  B {\bf 418}, 173 (1994)
  [arXiv:hep-th/9311062].

\bibitem{Sonnenschein:1988ug}
  J.~Sonnenschein,
  ``Chiral bosons,''
  Nucl.\ Phys.\  B {\bf 309}, 752 (1988).


\bibitem{Dabholkar:2005ve} A.~Dabholkar and C.~Hull, ``Generalised T-duality and non-geometric backgrounds,'' JHEP {\bf 0605} (2006) 009 [arXiv:hep-th/0512005].


\bibitem{ICBandHZ}
S.~Avramis, J.P.~Derendinger and N.~Prezas,
in progress.


\end{thebibliography}

\end{document}